\documentclass[onecolumn,draftclsnofoot,12pt]{IEEEtran}

\usepackage{amsfonts}
\usepackage{times}
\usepackage{graphicx}
\usepackage{latexsym}
\usepackage{dsfont}
\usepackage{amssymb}
\usepackage{amsmath}
\usepackage{cite}
\usepackage{verbatim}
\usepackage{subfigure}
\newcommand{\figref}[1]{{Fig.}~\ref{#1}}

\def\bb0{{\mathbb{0}}}

\def\ba{{\mathbf{a}}}
\def\bb{{\mathbf{b}}}
\def\bc{{\mathbf{c}}}

\def\bf{{\mathbf{f}}}
\def\bg{{\mathbf{g}}}
\def\bh{{\mathbf{h}}}

\def\bn{{\mathbf{n}}}

\def\br{{\mathbf{r}}}

\def\bv{{\mathbf{v}}}

\def\bx{{\mathbf{x}}}
\def\by{{\mathbf{y}}}
\def\bz{{\mathbf{z}}}
\def\b0{{\mathbf{0}}}

\def\bA{{\mathbf{A}}}
\def\bB{{\mathbf{B}}}

\def\bF{{\mathbf{F}}}
\def\bG{{\mathbf{G}}}
\def\bH{{\mathbf{H}}}
\def\bI{{\mathbf{I}}}

\def\bN{{\mathbf{N}}}

\def\bP{{\mathbf{P}}}
\def\bQ{{\mathbf{Q}}}
\def\bR{{\mathbf{R}}}

\def\bU{{\mathbf{U}}}
\def\bV{{\mathbf{V}}}
\def\bW{{\mathbf{W}}}
\def\bX{{\mathbf{X}}}
\def\bY{{\mathbf{Y}}}

\def\cA{\mathcal{A}}
\def\cB{\mathcal{B}}

\def\cN{\mathcal{N}}

\def\cS{\mathcal{S}}

\def\sf0{{\mathsf{0}}}

\def\Nt{{N_t}}

\allowdisplaybreaks

\usepackage{latexsym}
\usepackage{graphicx}
\usepackage{amsmath}
\usepackage{amssymb}
\usepackage{amsfonts}
\usepackage{epsfig}
\usepackage{cite}   
\usepackage{calc}
\usepackage{color}
\usepackage{theorem}
\usepackage{url}
\usepackage{subfigure}
\usepackage{cases}
\usepackage{verbatim}
\usepackage{array}
\usepackage{dcolumn}
\usepackage{multirow}
\usepackage{tabularx}
\usepackage{setspace}
\usepackage{algorithm}
\usepackage{algcompatible}
\usepackage{upgreek}

\usepackage{multicol}

\usepackage{balance}

\newtheorem{thm}{Theorem}

\newtheorem{prop}{Proposition}
\newtheorem{defn}{Definition}

\newcommand{\sref}[1]{{Section}~\ref{#1}}
\newcommand{\algoref}[1]{{Algorithm}~\ref{#1}}
\newcommand{\thmref}[1]{{Theorem}~\ref{#1}}
\newcommand{\propref}[1]{{Proposition}~\ref{#1}}
\newcommand{\bpf}{\begin{proof}}
\newcommand{\epf}{\end{proof}}
\newcommand{\bdefn}{\begin{defn}}
\newcommand{\edefn}{\end{defn}}

\usepackage{color}

\newcommand{\pinv}[1]{\ensuremath{#1^{\dagger}}} 	

\newenvironment{proof}{ \textbf{Proof:} }{ \hfill $\Box$}

\def\bW0{{\mathbb{0}}}

\def\ba{{\mathbf{a}}}
\def\bW{{\mathbf{b}}}
\def\bc{{\mathbf{c}}}

\def\bff{{\mathbf{f}}}
\def\bg{{\mathbf{g}}}
\def\bh{{\mathbf{h}}}

\def\bn{{\mathbf{n}}}

\def\br{{\mathbf{r}}}

\def\bv{{\mathbf{v}}}

\def\bx{{\mathbf{x}}}
\def\by{{\mathbf{y}}}
\def\bz{{\mathbf{z}}}
\def\b0{{\mathbf{0}}}

\def\bA{{\mathbf{A}}}
\def\bB{{\mathbf{B}}}

\def\bF{{\mathbf{F}}}
\def\bG{{\mathbf{G}}}
\def\bH{{\mathbf{H}}}
\def\bI{{\mathbf{I}}}

\def\bN{{\mathbf{N}}}

\def\bP{{\mathbf{P}}}
\def\bQ{{\mathbf{Q}}}
\def\bR{{\mathbf{R}}}

\def\bU{{\mathbf{U}}}
\def\bV{{\mathbf{V}}}
\def\bW{{\mathbf{W}}}
\def\bX{{\mathbf{X}}}
\def\bY{{\mathbf{Y}}}

\def\cA{\mathcal{A}}
\def\cB{\mathcal{B}}

\def\cN{\mathcal{N}}

\def\cS{\mathcal{S}}

\def\sf0{{\mathsf{0}}}

\def\Nt{N}
\def\Nrf{M}
\def\NantT{N_\mathrm{T}}
\def\NrfT{M_\mathrm{T}}
\def\DT{D_\mathrm{T}}
\def\NantR{N_\mathrm{R}}
\def\NrfR{M_\mathrm{R}}
\def\DR{D_\mathrm{R}}

\def \bAdt {\bA_{\rm{T}}}
\def \bAdr {\bA_{\rm{R}}}

\def \bGdt {\bG_{t}}

\begin{document}
\title{Spatial Channel Covariance Estimation \\ for the Hybrid MIMO Architecture: \\ 
A Compressive Sensing Based Approach}
\author{
Sungwoo~Park 
and Robert~W. Heath Jr.
\thanks{S. Park  and R. W. Heath Jr. are with the Wireless Networking and Communication Group (WNCG), Department of Electrical and Computer Engineering, The University of Texas at Austin, TX, 78701 USA. (e-mail: \{swpark96, rheath\}@utexas.edu).  }
\thanks{This work is supported in part by the National Science Foundation under Grant No. 1711702, and by a gift from Huawei Technologies.}
}
\maketitle

\begin{abstract} 
Spatial channel covariance information can replace full knowledge of the entire channel matrix  for designing analog precoders in hybrid multiple-input-multiple-output (MIMO) architecture.
Spatial channel covariance estimation, however, is challenging for the hybrid MIMO architecture because the estimator operating at baseband can only obtain a lower dimensional  pre-combined signal through fewer radio frequency (RF) chains than antennas. 
In this paper, we propose two approaches for covariance estimation  based on compressive sensing techniques.
One is to apply a time-varying sensing matrix, and the other is to exploit the prior knowledge that the covariance matrix is Hermitian. We present the rationale of the two ideas and validate the superiority of the proposed methods by  theoretical analysis  and numerical simulations.
We conclude the paper by extending the proposed algorithms from narrowband massive MIMO systems with a single receive antenna to wideband systems with multiple receive antennas.  
 \end{abstract}

\section{Introduction}\label{sec:intro}

One  way to increase coverage and capacity in wireless communication systems is to use a large number of antennas. 
For instance, millimeter wave systems use large antenna arrays to obtain high array gain,   thereby increasing cellular coverage \cite{Roh2014,HeathJr2015}. Sub-6 GHz systems are also likely to equip many antennas at a base station (BS) to increase cellular spectral efficiency by transmitting data to many users simultaneously via massive MIMO systems \cite{Marzetta2010,Hoydis2013}.
Using a large number of antennas in a conventional MIMO architecture, however, results in high cost and high power consumption because each antenna requires its own RF chain \cite{MolischCommMag2017}.
To solve this problem, 
hybrid analog/digital precoding reduces the number of RF chains by dividing the linear process between the analog RF part  and the  digital baseband part.
Several hybrid precoding techniques have been proposed for single-user MIMO \cite{ElAyach2014,Alkhateeb2013,Yu2016,Park2017twc} and multi-user MIMO \cite{Ni2016,Sohrabi2016,Alkhateeb2015,Adhikary2013,Park2017tsp} in either sub-6 GHz systems or millimeter wave systems.

Most prior work on hybrid precoding assumes that full channel state information at the transmitter (CSIT) for all antennas is available when designing the analog precoder. 
Full CSIT, however,  is difficult to obtain even in time-division duplexing (TDD) systems if many antennas are employed. Furthermore, the hybrid structure makes it more difficult to estimate the full CSIT of the entire channel matrix for all antennas because the estimator can only see a lower dimensional representation of the entire channel due to the reduced number of RF chains.
Although some channel estimation techniques for the hybrid architecture were proposed   by using compressive sensing techniques  that exploit spatial channel sparsity \cite{Alkhateeb2014JSTSP,Lee2016,SuryaprakashICASSP2016},
these techniques assume that  channel does not change during  the estimation process which requires many measurements over time. Consequently, these channel estimation techniques can be applied only to  slowly varying channels. 
 
Unlike the hybrid precoding techniques that require full CSIT, there exists another type of hybrid precoding that  uses long-term channel statistics such as spatial channel covariance  instead of full CSIT in the analog precoder design \cite{Adhikary2013,Park2017tsp,LiTWC2017,RialCAMSAP2015}. This approach has advantages over using full CSIT. 
First, the spatial channel covariance  is well modeled as constant over many channel coherence intervals  \cite{Love2008,LiTWC2017}. 
Second, the spatial channel covariance is constant over frequency in general \cite{Bjornson2009}, so covariance information is suitable for the wideband hybrid precoder design problem where one common analog precoder must be shared by all subcarriers in wideband OFDM systems \cite{Park2017twc}. For these reasons,  spatial channel covariance information is a promising alternative to full CSIT for hybrid MIMO architecture. 

Although several techniques have been developed to estimate spatial channel covariance or  its simplified version such as subspace or angle-of-arrival (AoA)   for hybrid architecture,   these methods have  practical issues.
For example, 
the sparse array developed in \cite{ShakeriArianadaLeus2012::2} and the coprime sampling used in \cite{RomeroLeus2013::11} reduce the number of RF chains for the covariance or AoA estimation as they disregard some of the antennas by exploiting  redundancy in linear arrays of equidistant antennas. These methods, however, 
have a limitation on the configuration of the number of RF chains and antennas. In \cite{DBLP:journals/corr/HaghighatshoarC15::9}, various estimation methods  were proposed based on  convex optimization problems coupled with the coprime sampling. The proposed methods, though,  require many iterations to converge.  
A subspace estimation method using the Arnoldi iteration was proposed for millimeter wave systems  in \cite{GhauchJSTSP2016}. 
The estimation target of this method is not the subspace of a spatial channel covariance matrix but that of a channel matrix itself, and  the channel matrix must be constant over time during the estimation process. Consequently,  it is difficult to apply the method in \cite{GhauchJSTSP2016} to time-varying channels.  
The subspace estimation method proposed in \cite{ChiEldarCalderbank2013::5}  does not use any symmetry properties of the covariance matrix, while the AoA estimation for the hybrid structure in \cite{ChuangWuLie2015::12}  does not exploit the sparse property of millimeter wave channels. In \cite{PengCL2015,LeeTC2016},  AoA estimation methods for the hybrid structure were proposed by applying compressive sensing techniques with vectorization but do not fully exploit the slow varying AoA property. 
Instead of using a vector-type compressive sensing, matrix-type compressive sensing techniques were developed for so-called multiple measurement vector (MMV) problems \cite{CotterRaoEnganDelgado2005::6, ChenTSP2006, DuarteTSP2011,Determe2016}.  
Most prior work on the MMV problems, however, assumes that the  sensing matrix is fixed over time to model the problem in a matrix form, which is not an efficient strategy when the measurement size becomes as small as the sparsity level.

In this paper, we propose a novel spatial channel covariance estimation technique for the hybrid MIMO architecture. Considering spatially sparse channels, we develop the estimation technique based on compressive sensing techniques that leverage the channel sparsity. 
Between two different approaches in the compressive sensing field (convex optimization algorithms vs. greedy algorithms), we focus on the greedy approach because they provide a similar performance  to convex optimization algorithms in spite of their  simpler  complexity \cite{Book::CompressedSensing_Eldar}.  
Based on well-known greedy algorithms such as orthogonal matching pursuit (OMP) and simultaneous OMP (SOMP), we improve  performance by applying two key ideas: one is to use a time-varying sensing matrix, and the other is to exploit the fact that covariance matrices are Hermitian. Motivated by the fact that SOMP is a generalized version of OMP, we develop a further generalized version of SOMP by  incorporating the proposed ideas. The algorithm modification is simple, but the performance improvement is significant, 
which is demonstrated by numerical and analytical results.

We first develop the spatial channel covariance estimation work for a simple scenario where a mobile station (MS) has a single antenna in narrowband systems. 
Preliminary results were presented in the conference version of this paper \cite{ParkAsilomar2016}. In this paper,  we  add two new contributions to our prior work. First, we present theoretical analysis to validate the rationale of the two proposed ideas and show the superiority of the proposed algorithm. The theoretical  analysis demonstrates that the use of a time-varying sensing matrix  dramatically improves the covariance estimation performance, in particular, when the number of RF chains is not so large and similar to the number of channel paths. The analytical results also disclose that exploiting the Hermitian property of the covariance matrix provides an additional gain.  
Second, we extend the estimation work to other scenarios. Considering that an MS as well as a BS has hybrid architecture with multiple antennas and RF chains, we modify the proposed algorithm to adapt to the situation where analog precoders as well as analog  combiners change over time. We also modify the algorithm for the wideband systems 
by using the fact that  frequency selective baseband combiners  do  not improve the estimation performance.

We use the following notation throughout this paper: $\bA$ is a matrix, $\ba$ is a vector, $a$ is a scalar, and $\mathcal{A}$ is a set.
$\|\ba \|_0$ and $\|\ba \|_2$ are the  $l_0$-norm  and $l_2$-norm of a vector $\ba$. 
 $\|\bA \|_F$ denotes a Frobenius norm.
$\bA^T$, $\bA^C$,  $\bA^*$, and $\pinv{\bA}$ are  transpose, conjugate,  conjugate transpose, and Moore-Penrose pseudoinverse. 
$[\bA]_{i,:}$ and $[\bA]_{:,j}$ are the $i$-th row and $j$-th column of the matrix $\bA$.
$[\bA]_{:,\cS}$ denotes a matrix whose columns are composed of $[\bA]_{:,j}$ for $j \in \cS$. 
If there is nothing ambiguous in the context,  $[\bA]_{:,j}$, $[\bA]_{:,\cS}$, $[\bA_t]_{:,j}$, and $[\bA_t]_{:,\cS}$ are  replaced by  $\ba_j$, $\bA_{\cS}$, $\ba_{t,j}$, and $\bA_{t,\cS}$.
$\cA \setminus \cB$ is  the relative complement of $\cA$ in $\cB$,   
$\mathrm{diag}(\bA)$ is a column vector whose elements are the diagonal elements of $\bA$, and $|\cA|$ is the cardinality of $\cA$.
$\bA \otimes \bB$, $\bA \circledcirc \bB$, and  $\bA \odot \bB$ denote the Kronecker product, the Hadamard product, and  the column-wise Khatri-Rao product.   
  $\bA \stackrel{(K)}{\odot} \bB$ is the generalized Khatri-Rao product with respect to $K$ partitions, which is defined as $\bA \stackrel{(K)}{\odot} \bB = \begin{bmatrix}  \bA_1 \otimes \bB_1 & \cdots &  \bA_K \otimes \bB_K \end{bmatrix}$
where $\bA = \begin{bmatrix}  \bA_1  & \cdots &  \bA_K  \end{bmatrix}$ and $\bB = \begin{bmatrix}  \bB_1  & \cdots &  \bB_K  \end{bmatrix}$.

\section{System model and preliminaires:  covariance estimation via compressive sensing based channel estimation}\label{sec:sys_model}
In this section, we present the  system model  and briefly overview  prior work on compressive sensing based channel estimation followed by covariance calculation. 
\subsection{System model}\label{subsec:sys_model}

Consider a  system where a BS with $\Nt$ antennas and $\Nrf$ RF chains communicates with an MS. We focus on a narrowband massive MIMO system where an MS with a single antenna; we extend the work to the multiple-MS-antenna case and the wideband case in \sref{sec:extension_to_others}. 
Let $L$ be the number of channel paths, $g_{\ell,t}$ be a time-varying channel coefficient of the $\ell$-th channel path at the $t$-th snapshot, and $\ba(\phi_{\ell})$ be an array response vector associated with the AoA of the $\ell$-th channel path $\phi_{\ell}$. 
Assuming that AoAs do not change during $T$ snapshots, the uplink channel at the $t$-th snapshot can be represented as
\begin{equation}\label{eq:ch_model_org}
\bh_t = \sum_{\ell=1}^{L} g_{\ell,t} \ba(\phi_{\ell}). 
\end{equation}
Considering spatially sparse channels,  
the channel model in  \eqref{eq:ch_model_org} can be approximated in the compressive sensing framework as 
\begin{equation}\label{eq:ch_model_CS}
\bh_t = \bA \bg_t  ,
\end{equation}
where $\bA \in \mathbb{C}^{\Nt \times D} $ is an  dictionary matrix whose $D$ columns are composed of the array response vectors associated with a predefined set of AoAs, and $\bg_t  \in \mathbb{C}^{D \times 1}$ is a sparse vector with only $L$ nonzero elements whose positions and values correspond to their AoAs and path gains. 

Let $d_t$ be a training symbol with $|d_t|=1$, and $\bz_t$ be a Gaussian noise with $\mathcal{CN}(\mathbf{0},\sigma^2 \bI)$. The received signal can be represented as
\begin{align}\label{eq:rx_model_org}
\bx_t = \bA  \bg_t d_t + \bz_t.
\end{align}
Combined with an analog combining matrix  $\bW_t \in \mathbb{C}^{\Nrf \times \Nt}$, the received signal  multiplied by $d_t^*$ at baseband is expressed as
\begin{align}\label{eq:rx_bb_model_vec}
\by_t =  d_t^* \bW_t \bx_t  =  \mathbf{\Phi}_t \bg_t + \bW_t \bn_t,  
\end{align}
where 
$ \mathbf{\Phi}_t  =  \bW_t \bA$ denotes an overall sensing matrix and $\bn_t = d_t^* \bz_t \sim\mathcal{CN}(\mathbf{0},\sigma^2 \bI)$.
By letting $\bR_{\bg} = \mathbb{E}[ \bg_t\bg^*_t ] $, the spatial channel covariance matrix is expressed as 
$\bR_{\bh} = \mathbb{E}[ \bh_t\bh^*_t ] = \bA \bR_{\bg}  \bA^*$.
The goal is to estimate $\bR_{\bh} $ with given  $\by_1, ..., \by_T$.

\subsection{Preliminaires:  covariance estimation via  compressive sensing based channel estimation}\label{subsec:using_prior_work}

Some prior work developed compressive channel estimators for the hybrid architecture \cite{Alkhateeb2014JSTSP,Lee2016,SuryaprakashICASSP2016}. Once the channel vectors are estimated, the covariance can be  calculated from the estimated channels. In this section, we overview two estimation approaches to make a comparison.

A  baseline technique to estimate the channel vector at each time  is formulated as
\begin{equation}\label{eq:SMVperSnapshot_criterion}
\begin{split}
\min_{ \bg_t }  \| \by_t  - \mathbf{\Phi}_t\bg_t  \|_2   \;\;
{\rm{s.t.}} \;\;  \| \bg_t   \|_0 \leq L .
\end{split}  
\end{equation}
\noindent This optimization problem is known as a single measurement vector (SMV) problem and can be solved by the  OMP  algorithm described  in \algoref{alg:OMP}, which
 is a simple but efficient approach among various solutions to the SMV problem \cite{DBLP:journals/corr/Mendez-RialRPAH15::10}. 
Once the $ \bg_t $'s are estimated during $T$ snapshots, then the  covariance matrix of the channel can be calculated as
\begin{figure}[t]
\begin{minipage}[t]{1.00\textwidth}
\setstretch{1.0}
\begin{multicols}{2}
{
\begin{minipage}{.45\textwidth}
\begin{algorithm}[H]
\caption{OMP}
\label{alg:OMP}
\begin{algorithmic}
\State Input:  $\mathbf{\Phi}, \by,$ and $L$
\State Initialize: $\bv=\by,   \mathcal{S}=\O, \hat{\bg}=\mathbf{0}$
\FOR {$n =1:L $}
\State $j = \arg \max_{i} |  \boldsymbol{\phi}_{i}^* \bv |$
\State $\mathcal{S} = \mathcal{S}  \cup \{ j \}$
\State $\bv = \left( \bI - \mathbf{\Phi}_{\mathcal{S}} \mathbf{\Phi}_{\mathcal{S}}^{\dagger}  \right) \by $
\ENDFOR
\State $\hat{\bg}_{\mathcal{S}} = \mathbf{\Phi}^{\dagger}_{\mathcal{S}}    \by$
\State Output: $\hat{\bg}$ 
\end{algorithmic}
\end{algorithm}
\end{minipage}
}
{
\begin{minipage}{.45\textwidth}
\begin{algorithm}[H]
\caption{SOMP}
\label{alg:SOMP}
\begin{algorithmic}
\State Input:  $\mathbf{\Phi}, \bY,$ and $L$
\State Initialize: $\bV=\bY,   \mathcal{S}=\O, \hat{\bG}=\mathbf{0}$
\FOR {$n =1:L $}
\State $j = \arg \max_{i} \|\boldsymbol{\phi}_{i}^*\bV \|_2$
\State $\mathcal{S} = \mathcal{S}  \cup \{ j \}$
\State $\bV = \left( \bI - \mathbf{\Phi}_{\mathcal{S}} \mathbf{\Phi}^{\dagger}_{\mathcal{S}} \right) \bY $
\ENDFOR
\State $[\hat{\bG}]_{\mathcal{S},:} = \mathbf{\Phi}^{\dagger}_{\mathcal{S}}   \bY$
\State Output: $\hat{\bG}$ 
\end{algorithmic}
\end{algorithm}
\end{minipage}
}
\end{multicols}
\end{minipage}
\end{figure}
\begin{equation}\label{eq:SMVperSnapshot_Rh}
\hat{\bR}_\bh = \bA \left( \frac{1}{T} \sum_{t=1}^{T}\bg_t \bg^*_t \right) \bA^* .
\end{equation}

In contrast to  SMV, another  compressive sensing technique called MMV \cite{CotterRaoEnganDelgado2005::6} 
exploits the observation that  $\bg_1, ..., \bg_T$ shares a common support if the AoAs do not change during the estimation process. The  received signal model in \eqref{eq:rx_bb_model_vec} can be written in a matrix form as
 \begin{align}\label{eq:rx_bb_model_mat}
\bY  = \mathbf{\Phi} \bG + \bW \bN,  
\end{align}
where $\bY = \begin{bmatrix} \by_1 & \cdots & \by_T \end{bmatrix}$,  $\bG = \begin{bmatrix} \bg_1 & \cdots & \bg_T \end{bmatrix}$, and $\bN = \begin{bmatrix} \bn_1 & \cdots & \bn_T \end{bmatrix}$. 
Note that in this MMV problem format, the sensing matrix $\mathbf{\Phi}$ must be  common over time. 
The optimization problem in the MMV is formulated as
\begin{equation}\label{eq:MMV_criterion}
\begin{split} 
\min_{\bG} \|  \bY  -   \mathbf{\Phi}  \bG   \|_F  \;\;
{\rm{s.t.}} \;\;  \| \bG   \|_{{\rm{row}},0} \leq L, 
\end{split} 
\end{equation}
where $  \|  \bG\|_{{\rm{row}},0} $ represents the row sparsity defined as 
$ \|  \bG\|_{{\rm{row}},0} = \left | \bigcup_t {\rm{supp}} \left( [\bG]_{:,t} \right) \right |$.
This optimization problem can be solved by the SOMP algorithm 
described in \algoref{alg:SOMP} 
\cite{TroppICASSP05,DBLP:journals/corr/DetermeLJH15:13}. 
Regarding the selection rule of SOMP, the $\ell_2$-norm in \algoref{alg:SOMP}  can be replaced by the $\ell_1$-norm because there is no significant difference between  the two in terms of performance \cite{ChenTSP2006}.
 We use  $\ell_2$-norm throughout this paper for analytical tractability.
Although SOMP is known to outperform OMP in general,  SOMP still needs improvement, particularly when the number of RF chains is small, as will be discussed in \sref{subsec:proposed_timevarying_analog_matrix}.

\section{Covariance estimation for hybrid architecture: two key ideas}\label{sec:two_main_ideas}

In this section, we present two key ideas to improve covariance estimation work based on compressive sensing techniques. First, we develop an estimation algorithm by applying a time-varying sensing matrix. Second, we propose another algorithm that exploits the Hermitian property of the covariance matrix. Finally, we combine the two proposed algorithms. 

\subsection{Applying a time-varying analog combining matrix}\label{subsec:proposed_timevarying_analog_matrix}

Although compressive sensing can reduce the required number of measurements, there is a limitation on reducing the measurement size. This lower bound is known to be $\Nrf = \mathcal{O}\hspace{-2pt}\left(L\log\left(\frac{D}{L}\right)\right) $ for both  SMV and  MMV when the sensing matrix satisfies the restricted isometry property (RIP) condition.  \cite{DavenportWakin2010}. 
One possible solution to overcome this limitation is to extend the measurement vector size into the time domain, i.e., gathering the measurement vectors of multiple snapshots with different sensing matrices over time. The same approach is also used in \cite{Alkhateeb2014JSTSP} and \cite{DBLP:journals/corr/Mendez-RialRPAH15::10}, 
where the 
measurements are stacked together as
\begin{equation}\label{eq:timevarying_timeslot_static}
\begin{bmatrix} \by_1 \\ \vdots \\ \by_T \end{bmatrix} = \begin{bmatrix} \bW_1 \bA \\ \vdots \\ \bW_T \bA \end{bmatrix} \bg + \begin{bmatrix} \bW_1 \bn_1 \\ \vdots \\ \bW_T \bn_T \end{bmatrix}  = \begin{bmatrix} \mathbf{\Phi}_1  \\ \vdots \\ \mathbf{\Phi}_T \end{bmatrix} \bg + \tilde{\bn}.
\end{equation}
Note that the key assumption in \eqref{eq:timevarying_timeslot_static} is that  $\bg$ is constant during the estimation process. Consequently, this technique can be applied only to static channels, not time-varying channels. 

It is worthwhile to compare the MMV signal model in \eqref{eq:rx_bb_model_mat} with the signal model where the time-varying combining matrix is used for the static channel as shown in \eqref{eq:timevarying_timeslot_static}. In \eqref{eq:rx_bb_model_mat}, $\by_t$'s are stacked in columns due to the fact that $\bg_t$ changes over time but  $\mathbf{\Phi}$ is fixed. In contrast,  the signal model in \eqref{eq:timevarying_timeslot_static} adopts a row-wise stack of $\by_t$ because $\mathbf{\Phi}_t$ changes over time but $\bg$ is fixed. If both $\mathbf{\Phi}_t$ and $\bg_t$ change over time, we can not stack $\by_t$'s in either columns or rows and thus need another approach. 
Regarding this scenario, two questions arise: 1) is it useful to employ a time-varying sensing matrix for a time-varying channel? 2) how can we recover the data in this time-varying sensing matrix and time-varying data?
We will show that the time-varying sensing matrix can increase the  recovery success rate especially when $M$ is not much larger than $L$ by analysis in \sref{sec:theoretical_analysis} and  simulations in \sref{sec:sim_results}. In this subsection, we develop a recovery algorithm to answer the problem of recovery with a time-varying sensing matrix.  

To apply the time-varying matrix concept to  SOMP, we first focus on the fact that SOMP is a generalized version of  OMP. 
The reason is that  the optimization problem of MMV in \eqref{eq:MMV_criterion} can be rewritten as
\begin{equation}\label{eq:MMV_criterion_alternative}
\begin{split} 
\min_{\bg_1,...,\bg_T}  \sum_{t=1}^T  \|  \by_t  -   \mathbf{\Phi}  \bg_t   \|_2^2  \;\;
{\rm{s.t.}} \;\;  \left | \bigcup_{t=1}^T {\rm{supp}} \left( \bg_t \right) \right |  \leq L.
\end{split} 
\end{equation}
Note that  SOMP is equivalent to  OMP when $T=1$. 
Unlike the original formulation in \eqref{eq:MMV_criterion}, this reformulated form gives an insight into how to apply the time-varying sensing matrix. 
We can further generalize the optimization problem in \eqref{eq:MMV_criterion_alternative} by replacing 
 $\mathbf{\Phi}$ with $\mathbf{\Phi}_t$ to reflect the time-varying feature of the sensing matrix. 
Noting that the selection rule of SOMP in \algoref{alg:SOMP} is equivalent to $j = \arg \max_{i} \|\boldsymbol{\phi}_{i}^*\bV \|_2^2 
=\arg \max_{i} \sum_{t=1}^T |\boldsymbol{\phi}_{i}^*\bv_t |^2$, we modify the selection rule  by replacing $\boldsymbol{\phi}_{i}$ with $\boldsymbol{\phi}_{t,i}(=[\mathbf{\Phi}_t]_{:,i})$ to adapt to the time-varying sensing matrix case. The residual matrix  in SOMP, $\bV = \left( \bI - \mathbf{\Phi}_{\mathcal{S}} \mathbf{\Phi}^{\dagger}_{\mathcal{S}} \right) \bY $,  can also be replaced by  $\bv_t = \left( \bI - \mathbf{\Phi}_{t,\mathcal{S}} \mathbf{\Phi}^{\dagger}_{t,\mathcal{S}} \right) \by_t $ for  $t=1,...,T$. The modified version of SOMP, which we call dynamic SOMP (DSOMP), is described in \algoref{alg:DSOMP}.

\subsection{Exploiting the Hermitian property of the covariance matrix}\label{subsec:using_covariance_structure}

The covariance estimation techniques using OMP, SOMP, and DSOMP in the previous subsections employ a two-step approach where the channel gain vectors $\bg_t$'s (and thus channel vectors $\bh_t$'s) are estimated and then the covariance matrix is calculated from  \eqref{eq:SMVperSnapshot_Rh}. If it is not the channel  but the covariance that needs to be estimated, the first step estimating the channel explicitly is unnecessary. In this subsection, we take a different approach that directly estimates  the covariance $\bR_{\bg}$   without  estimating the instantaneous channel gain vectors.

\begin{figure}[t]
	\centering
	\subfigure[center][{Unstructured}]{
		\includegraphics[width=.15\linewidth]{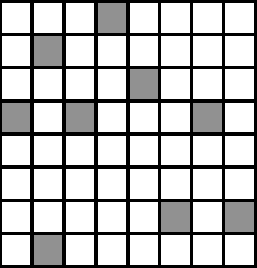}
		\label{fig:Fig_unstructured_sparse_matrix}}
	\subfigure[center][{Sparse rows}]{
		\includegraphics[width=.15\columnwidth]{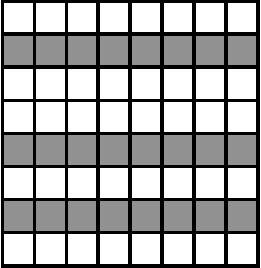}
		\label{fig:Fig_sparse_row_matrix}}
	\subfigure[center][{Sparse diagonal}]{
		\includegraphics[width=.15\columnwidth]{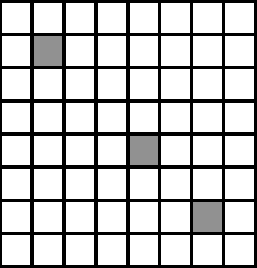}
		\label{fig:Fig_sparse_diagonal_matrix}}
	\subfigure[center][{Sparse Hermitian}]{
		\includegraphics[width=.15\columnwidth]{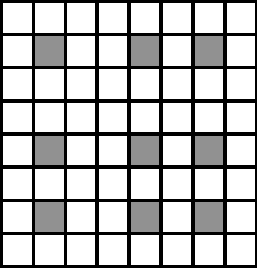}
		\label{fig:Fig_sparse_Hermitian_matrix}}
	\caption{Sparse matrix types: (a) an unstructured sparse matrix where the elements are randomly spread, (b) a structured sparse matrix with sparse rows, (c) a structured sparse diagonal matrix, and (d) a structured sparse Hermitian matrix.}
	\label{fig:Fig_Sparse_matrix_types}
\end{figure}

 The relationship between  $\bR_{\bg}$ and $\bR_{\by}$  is given by
\begin{equation}\label{eq:CMV_model_simple}
\bR_\by = \mathbf{\Phi} \bR_{\bg} \mathbf{\Phi}^* + \sigma^2  \bW \bW^*.
\end{equation}
If we consider a special case where $\bR_{\bg}$ is assumed to be a sparse diagonal matrix as in \figref{fig:Fig_sparse_diagonal_matrix}, 
the relationship between $\bR_{\bg}$ and $\bR_{\by}$ in \eqref{eq:CMV_model_simple}  can be rewritten as
\begin{equation}\label{eq:CMV_model_vec}
{\rm{vec}} (\bR_\by)  
 = \left( \mathbf{\Phi}^C \odot \mathbf{\Phi}  \right) \rm{diag}( \bR_{\bg} ) + \sigma^2{\rm{vec}} (\bW \bW^*), 
\end{equation}
by using vectorization and the column-wise Khatri-Rao product, then OMP can be directly applied to this reformulated problem without any modification  \cite{ArianandaLeus13}.
This approach, however, has a limitation in application to realistic scenarios because the covariance $\bR_{\bg}$ is  not a diagonal matrix in general \cite{ParkAsilomar2016}.  
Instead of assuming that $\bR_{\bg}$ is a sparse diagonal matrix, we consider $\bR_{\bg}$ to be a sparse Hermitian matrix as in \figref{fig:Fig_sparse_Hermitian_matrix}. 
As  MMV uses the fact that $\bG$ is a matrix with sparse rows as in \figref{fig:Fig_sparse_row_matrix}, we exploits the Hermitian property of the  covariance matrix $\bR_{\bg}$. 
The optimization problem is represented in a matrix form like MMV, 
\begin{equation}\label{eq:COMP_criterion}
\begin{split} 
\min_{\bR_{\bg}}  \|  \bR_{\bY}  - \mathbf{\Phi} \bR_{\bg}  \mathbf{\Phi} ^*    \|_F \;\;
{\rm{s.t.}} \;\;  \| \bR_{\bg}   \|_{{\rm{lattice}},0} \leq L, 
\end{split} 
\end{equation}
where $\| \bR_{\bg}  \|_{{\rm{lattice}},0}$ is defined as $ \|  \bR_{\bg}\|_{{\rm{lattice}},0} = \left| \bigcup_i {\rm{supp}} \left( [\bR_{\bg}]_{:,i} \right) \bigcup_j{\rm{supp}} \left( [\bR_{\bg}]_{j,:} \right) \right|$.

A greedy approach similar to OMP and SOMP can be applied to find a suboptimal solution. 
At each iteration, the algorithm needs to find the solution to the following sub-problems as 
\begin{equation}\label{eq:quad_min_sol}
\min_{\bR_{\bX}}  \|  \bR_{\bY}  - \mathbf{\Phi}_{\cS_n} \bR_{\bX}  \mathbf{\Phi} ^*_{\cS_n}    \|_F,
\end{equation}
where $\bR_{\bX}$ is a Hermitian matrix whose dimension is less than or equal to $L$. By using the least squares method along with vectorization, the optimal solution to \eqref{eq:quad_min_sol} is given by
\begin{equation}\label{eq:quad_min_sol_opt}
\bR_{\bX}^{\mathrm{(opt)}} = \mathbf{\Phi}_{\cS_n}^{\dagger}  \bR_{\bY} \left(\mathbf{\Phi}_{\cS_n}^{\dagger}\right)^*,
\end{equation}
where $ \mathbf{\Phi}_{\cS_n}^{\dagger}=\left(\mathbf{\Phi}_{\cS_n}^* \mathbf{\Phi}_{\cS_n}   \right)^{-1} \mathbf{\Phi}_{\cS_n}^*$.
Note that if $\bR_{\bY}$ is Hermitian, $\bR_{\bX}^{\mathrm{(opt)}}$ is also  Hermitian.
The proposed algorithm  using \eqref{eq:quad_min_sol_opt}, which we call covariance OMP (COMP), is described in \algoref{alg:COMP}. 
Note that the proposed COMP uses quadratic forms instead of linear forms.

The time-varying sensing matrix concept in \sref{subsec:proposed_timevarying_analog_matrix} can also be applied to COMP. The generalization of COMP, however, requires a careful modification. In  covariance estimation based on COMP, the calculation of $\hat{\bR}_\by$ is followed by the estimation of $\hat{\bR}_{\bg}$. This is because $\hat{\bR}_\by$ can be represented as a function of $\hat{\bR}_{\bg}$ as 
\begin{equation}\label{eq:Relation_btw_Ry_Rg}
\hat{\bR}_\by = \frac{1}{T} \sum_{t=1}^{T}\by_t \by_t ^* = \mathbf{\Phi} \left( \frac{1}{T} \sum_{t=1}^{T}\bg_t \bg_t ^* \right)  \mathbf{\Phi} ^* = \mathbf{\Phi} \hat{\bR}_{\bg}  \mathbf{\Phi} ^*.
\end{equation}
If the sensing matrix changes over time,  $\hat{\bR}_\by$ can not be represented as a function of $\hat{\bR}_{\bg}$ because
\begin{equation}\label{eq:Relation_btw_Ry_Rg_dynamic}
\hat{\bR}_\by =  \frac{1}{T} \sum_{t=1}^{T}\by_t \by_t ^* = \frac{1}{T} \sum_{t=1}^{T} \mathbf{\Phi}_t \bg_t \bg_t ^* \mathbf{\Phi}_t^* .
\end{equation}

\begin{figure}[t]
\begin{minipage}[t]{1.00\textwidth}
\setstretch{0.9}
\begin{multicols}{2}
{
\setstretch{0.8}
\begin{algorithm}[H]
\caption{Dynamic SOMP (DSOMP)}
\label{alg:DSOMP}
\begin{algorithmic}
\State Input:  $\mathbf{\Phi}_1,...,\mathbf{\Phi}_T, \bY, L$
\State Initialize: $\bV=\bY,   \mathcal{S}=\O, \hat{\bG}=\mathbf{0}$
\FOR {$n =1:L $}
\State $j = \arg \max_{i} \sum_{t=1}^{T} \left| \boldsymbol{\phi}_{t,i}^*  \bv_{t} \right|^2$
\State $\mathcal{S} = \mathcal{S}  \cup \{ j \}$
\State $\bv_t =\left(\bI - \mathbf{\Phi}_{t,\mathcal{S}} \mathbf{\Phi}^{\dagger}_{t,\mathcal{S}} \right) \by_t  , \forall t$
\ENDFOR
\State $[\hat{\bG}]_{\mathcal{S},:}=\begin{bmatrix} \mathbf{\Phi}^{\dagger} _{1,\mathcal{S}}  \by_{1}  & \cdots & \mathbf{\Phi}^{\dagger} _{T,\mathcal{S}}  \by_{T} \end{bmatrix} $
\State Output: $\hat{\bG}$
\end{algorithmic}
\end{algorithm}
}
{
\setstretch{0.85}
\begin{algorithm}[H]
\caption{Covariance OMP (COMP)}
\label{alg:COMP}
\begin{algorithmic}
\State Input:  $\mathbf{\Phi}, \bY, L$
\State Initialize: $\hspace{-2pt} \bV \hspace{-2pt} = \hspace{-2pt} \hat{\bR}_{\by} \hspace{-2pt} = \hspace{-2pt} \frac{1}{T} \bY \bY^* ,   \mathcal{S}=\O, \hat{\bR}_{\bg} = \mathbf{0}$
\FOR {$n =1:L $}
\State $j = \arg \max_{i}   \boldsymbol{\phi}_{i}^*\bV \boldsymbol{\phi}_{i} $
\State $\mathcal{S} = \mathcal{S}  \cup \{ j \}$
\State $ \bV =  \hat{\bR}_\by -  \mathbf{\Phi}_{\mathcal{S}} \mathbf{\Phi}^{\dagger}_{\mathcal{S}}    \hat{\bR}_{\by} \left( \mathbf{\Phi}_{\mathcal{S}} \mathbf{\Phi}^{\dagger}_{\mathcal{S}} \right)^{*} $ 
\ENDFOR
\State  $ [\hat{\bR}_{\bg}]_{\mathcal{S},\mathcal{S}}=\mathbf{\Phi}^{\dagger}_{\mathcal{S}}    \hat{\bR}_{\by} \left(\mathbf{\Phi}^{\dagger}_{\mathcal{S}} \right)^{*} $ 
\State Output: $ \hat{\bR}_{\bg}$
\end{algorithmic}
\end{algorithm}
}
\end{multicols}
\setstretch{0.8}
\begin{algorithm}[H]
\caption{Dynamic COMP (DCOMP)}
\label{alg:DCOMP}
\begin{algorithmic}
\State Input:  $\mathbf{\Phi}_1, ... , \mathbf{\Phi}_T,\bY , L$
\State Initialize: $\bV_t=\hat{\bR}_{\by,t} = \by_{t} \by_{t}^*, \forall t,   \mathcal{S}=\O, \hat{\bR}_{\bg}=\mathbf{0}$
\FOR {$n =1:L $}
\State $j = \arg \max_{i} \sum_{t=1}^{T}  \boldsymbol{\phi}_{t,i}^*\bV_t \boldsymbol{\phi}_{t,i} $
\State $\mathcal{S} = \mathcal{S}  \cup \{ j \}$
\State $\bV_t = \hat{\bR}_{\by,t} - \mathbf{\Phi}_{t,\mathcal{S}}  \mathbf{\Phi}^{\dagger}_{t,\mathcal{S}}  \hat{\bR}_{\by,t}  \left(\mathbf{\Phi}_{t,\mathcal{S}} \mathbf{\Phi}^{\dagger}_{t,\mathcal{S}} \right)^{*} ,  \forall t$
\ENDFOR
\State $[\hat{\bR}_{\bg}]_{\mathcal{S},\mathcal{S}} =  \frac{1}{T} \sum_{t=1}^T \mathbf{\Phi}^{\dagger}_{t,\mathcal{S}}  \hat{\bR}_{\by,t}  \left(\mathbf{\Phi}^{\dagger}_{t,\mathcal{S}} \right)^{*}$
\State Output: $\hat{\bR}_{\bg} $
\end{algorithmic}
\end{algorithm}
\end{minipage}
\end{figure}

For this reason, instead of using the sample covariance matrix of the measurement vectors $\hat{\bR}_\by$, we use a per-snapshot $\by_t \by_t^*$, which can be regarded as a one sample estimate of the sample covariance matrix. Note that 
this extreme sample covariance is not 
a diagonal matrix 
in general even when the channel paths are uncorrelated. 
Using the fact that $\by_t \by_t^*$ for all snapshots are sparse Hermitian matrices sharing the same positions of nonzero elements, we develop the dynamic COMP (DCOMP) from  COMP  in a similar way that we develop  DSOMP  from SOMP, which is described in \algoref{alg:DCOMP}. 
Note that  DCOMP becomes equivalent to COMP if $\mathbf{\Phi}_1 = \cdots =  \mathbf{\Phi}_T$.

\section{Theoretical analysis of the proposed alogirhtms}\label{sec:theoretical_analysis}

In this section, we analyze the benefit of the time-varying sensing matrix and compare SOMP, DSOMP, and DCOMP. In summary,  
the recovery success probability of SOMP, DSOMP, and DCOMP increases as the number of measurements $T$ increases.  The  recovery success probability of  SOMP, however,  saturates and does not approach one as $T$ goes to infinity even in the noiseless case if the number of RF chains $\Nrf$ is not much larger than the number of channel paths $L$. In contrast to  SOMP, 
the recovery success probability of  DSOMP and  DCOMP approaches one as $T$ increases for any $\Nrf$ and $L$. In other words,  DSOMP and  DCOMP can guarantee perfect recovery even when the number of RF chains is so small that both OMP and SOMP fail to recover the support even in the noiseless case.  
Finally, we will show that   DCOMP has a higher recovery success probability than  DSOMP.

Let us first consider how to design the sensing matrix $\mathbf{\Phi}$. For analytical tractability, we confine the dictionary size to $D=\Nt$ in this section. The algorithms, however, can be applied to more  general cases such as $D>\Nt$.  The sensing matrix can be represented as $\mathbf{\Phi} = \begin{bmatrix} \boldsymbol{\phi}_1& \cdots & \boldsymbol{\phi}_{\Nt} \end{bmatrix} = \bW  \bA$.
One possible option for the sensing matrix is to use a random analog precoding matrix $ \bW = \bW_{\mathrm{RF}}$ such that each element in $\bW_{\mathrm{RF}}$ has a constant amplitude and a random phase with  an independent uniform distribution in [0, $2\pi$]. Since the elements in the sensing matrix  designed in this way have a sub-Gaussian distribution, this simple random sensing matrix $\mathbf{\Phi}$ 
satisfies the RIP condition \cite{Book::CompressedSensing_Eldar}.
There are other desirable features that the sensing matrix needs to have. For example, it is desirable for the sensing matrix to have a small mutual coherence  defined as  $\rho = \max_{i \neq j} \frac{ |\boldsymbol{\phi}_j^* \boldsymbol{\phi}_i | }{\|\boldsymbol{\phi}_j \| \|\boldsymbol{\phi}_i \|}$ \cite{Book::CompressedSensing_Eldar}.
The mutual coherence 
has a lower bound $\sqrt{\frac{\Nt-\Nrf}{\Nrf(\Nt-1)}}$, which is known as Welch bound. This bound can be achievable if the column vectors in $\mathbf{\Phi}$ constitute an equal-norm equiangular tight frame, i.e., $\mathbf{\Phi}$ satisfies that 1) $\|\boldsymbol{\phi}_{i} \|=c_1, \forall i$, 2) $|\boldsymbol{\phi}_{j}^* \boldsymbol{\phi}_{i} |=c_2, \forall j \neq i$, and 3) $\mathbf{\Phi} \mathbf{\Phi}^* = c_3 \bI$. 
It is impossible for a sensing matrix  $\mathbf{\Phi}$ that is designed with a random phase analog combining matrix $\bW_{\mathrm{RF}}$ to meet these three conditions at all  times. Nevertheless, it is possible to meet the tight frame condition for any random analog precoding matrix $\bW_{\mathrm{RF}}$ if  the baseband combining matrix $\bW_{\mathrm{BB}}$ is employed  such that 
\begin{equation}\label{def_sensing_BB_mat}
\bW = \bW_{\mathrm{BB}}\bW_{\mathrm{RF}} = \left( \bW_{\mathrm{RF}}  \bW_{\mathrm{RF}} ^*\right) ^{-\frac{1}{2}} \bW_{\mathrm{RF}}.  
\end{equation} 
If the baseband combining matrix is designed as \eqref{def_sensing_BB_mat}, the columns in the overall sensing matrix $\mathbf{\Phi}$ constitute the tight frame because it satisfies 
\begin{equation}\label{sensing_tight_frame}
\begin{split}
\mathbf{\Phi} \mathbf{\Phi} ^* &= \bW_{\mathrm{BB}} \bW_{\mathrm{RF}} \bA \bA^* \bW_{\mathrm{RF}} ^* \bW_{\mathrm{BB}} ^*  
= \Nt \bI.
\end{split}
\end{equation}
Consequently, any  sensing matrix can be transformed to satisfy the tight frame condition by using  $ \bW_{\mathrm{BB}} = \left( \bW_{\mathrm{RF}}  \bW_{\mathrm{RF}} ^*\right) ^{-\frac{1}{2}} $.

The following theorem provides the exact condition for the perfect recovery condition at each iteration of the DSOMP algorithm. Note that  OMP and  SOMP are special cases of  DSOMP. 

\begin{thm}\label{thm:rho_def_DSOMP}
Let $\cS \subset \cN=\{1,2,...,N\}$ be an optimal support set with $|\cS| = L$ and $\bG$ be an ideal channel gain matrix.
Suppose that $\cS_o \subset \cS $ with $|\cS_o| = L_o$ was perfectly recovered  at the previous $L_o$ iterations  in the DSOMP algorithm. 
In the noiseless case, one of the elements in the optimal support $\cS$ can be perfectly recovered at the $(L_o+1)$-th iteration if and only if
\begin{equation}\label{def_rho_DSOMP0}
\begin{split}
\rho ^{\mathrm{(DS)}} (\cS, \cS_o, \mathbf{\Phi}_{1},...,\mathbf{\Phi}_{T},\bG)  
 &= \frac{ \max_{j \in \cN \setminus \cS}  \sum_{t=1}^{T} \left | \sum_{i \in \cS \setminus \cS_o} \boldsymbol{\psi}_{t,j}^* \boldsymbol{\psi}_{t,i}  {g}_{t,i} \right |^2 }   {\max_{j \in \cS \setminus \cS_o}  \sum_{t=1}^{T} \left | \sum_{i \in \cS \setminus \cS_o} \boldsymbol{\psi}_{t,j}^*   \boldsymbol{\psi}_{t,i}  {g}_{t,i} \right |^2 }  < 1,
\end{split}
\end{equation} 
where $\mathbf{\Psi}_t = \left( \bI - \mathbf{\Phi}_{t,\cS_o} \mathbf{\Phi}_{t,\cS_o}^{\dagger} \right) \mathbf{\Phi}_{t}$.
\end{thm}
\begin{IEEEproof}
See Appendix  \ref{app:proof_theorem_1}.
\end{IEEEproof}
Although the newly defined metric $\rho^{\mathrm{(DS)}}(\cS, \cS_o, \mathbf{\Phi}_{1},...,\mathbf{\Phi}_{T},\bG)$  identifies the exact condition for the perfect recovery condition, 
its dependence on the input data $\bG$ makes this metric less attractive. 
Some prior work  takes an approach to find  upper bounds of certain coherence-related metrics for any random input data \cite{TroppTit2004}, but the bounds are loose, which can be regarded as a worst case analysis. 
In addition, those upper bounds are meaningful only for the case where $\Nrf$ is significantly larger than $L$. In the extreme case where $\Nrf$ and $L$ are similar in value, the upper bounds are too loose to give a useful insight. 

Instead of an upper bound, we focus on the distribution of   $\rho^{\mathrm{(DS)}}(\cS, \cS_o, \mathbf{\Phi}_{1},...,\mathbf{\Phi}_{T},\bG)$ considering   $\cS$, $\cS_o$, $ \mathbf{\Phi}_{1},...,\mathbf{\Phi}_{T}$, and $\bG$ as random variables. 
While the exact distribution 
depends on the distribution of $\bG$, 
\propref{prop:SOMP_rho} and \thmref{thm:DSOMP_convergence}  show that $\rho^{\mathrm{(DS)}}(\cS, \cS_o, \mathbf{\Phi}_{1},...,\mathbf{\Phi}_{T},\bG)$ converges to a value that does not depend on $\bG$ as $T$ becomes larger.

\begin{prop} \label{prop:SOMP_rho}
Suppose that  ${g}_{t,i}$ are independent random variables with zero mean and unit variance. Let $\rho^{\mathrm{(S)}}(\cS, \cS_o, \mathbf{\Phi},\bG)$ denote $\rho^{\mathrm{(DS)}}(\cS, \cS_o, \mathbf{\Phi}_1 ,...,\mathbf{\Phi}_T,\bG)$ in the SOMP case where $\mathbf{\Phi}_1 = \cdots = \mathbf{\Phi}_T = \mathbf{\Phi}$. As $T \rightarrow \infty$, $\rho^{\mathrm{(S)}}(\cS, \cS_o, \mathbf{\Phi},\bG)$ converges to
\begin{equation}\label{def_rho_SOMP}
\rho^{\mathrm{(S)}}(\cS, \cS_o, \mathbf{\Phi},\bG) \rightarrow\rho^{\mathrm{(S)}}_{T \rightarrow \infty}(\cS, \cS_o, \mathbf{\Phi})  = \frac{ \max_{j \in \cN \setminus \cS}   \sum_{i \in \cS \setminus \cS_o} \left |  \boldsymbol{\psi}_{j}^* \boldsymbol{\psi}_{i}   \right |^2 }   {\max_{j \in \cS \setminus \cS_o}  \sum_{i \in \cS \setminus \cS_o} \left |  \boldsymbol{\psi}_{j}^*   \boldsymbol{\psi}_{i}   \right |^2 },  
\end{equation}
where $\mathbf{\Psi} =  \left( \bI - \mathbf{\Phi}_{\cS_o} \mathbf{\Phi}_{\cS_o}^{\dagger} \right) \mathbf{\Phi}$.
\end{prop}
\begin{IEEEproof}
See Appendix  \ref{app:proof_prop_1}.
\end{IEEEproof}

Note that $\rho^{\mathrm{(S)}}_{T \rightarrow \infty}(\cS, \cS_o, \mathbf{\Phi}) $ in \eqref{def_rho_SOMP} can be rewritten as 
\begin{equation}\label{def_rho_SOMP_re1}
\begin{split}
\rho^{\mathrm{(S)}}_{T \rightarrow \infty}(\cS, \cS_o, \mathbf{\Phi}) 
&= \frac{ \max_{j \in \cN \setminus \cS} \left(   \sum_{i \in \cS \setminus \cS_o}\left | \boldsymbol{\psi}_{j}^* \boldsymbol{\psi}_{i}   \right |^2  \right)}   {\max_{j \in \cS \setminus \cS_o}    \left( |\boldsymbol{\psi}_{j}|^4 + \sum_{\substack{i \neq j \\ i \in \cS \setminus \cS_o}} \left | \boldsymbol{\psi}_{j}^*   \boldsymbol{\psi}_{i}   \right |^2 \right)}.
\end{split}
\end{equation}
In the OMP case, $\rho^{\mathrm{(DS)}}(\cS, \cS_o, \mathbf{\Phi}_{1},...,\mathbf{\Phi}_{T},\bG)$  becomes equivalent to
\begin{equation}\label{def_rho_OMP}
\begin{split}
\rho^{\mathrm{(O)}}(\cS, \cS_o, \mathbf{\Phi},\bg)  
&= \frac{ \max_{j \in \cN \setminus \cS} \left(   \sum_{i \in \cS \setminus \cS_o}\left | \boldsymbol{\psi}_{j}^* \boldsymbol{\psi}_{i}   \right |^2  + X_j \right)}   {\max_{j \in \cS \setminus \cS_o}    \left( |\boldsymbol{\psi}_{j}|^4 + \sum_{\substack{i \neq j \\ i \in \cS \setminus \cS_o} } \left | \boldsymbol{\psi}_{j}^*   \boldsymbol{\psi}_{i}   \right |^2 + X_j \right)},
\end{split}
\end{equation}
where $X_j =   \sum_{i_1 \in \cS \setminus \cS_o} \sum_{\substack{ i_2 \in \cS \setminus \cS_o \\ i_2 \neq  i_1 } }  {g}_{i_1} {g}_{i_2}^* \boldsymbol{\psi}_{j}^* \boldsymbol{\psi}_{i_1} \boldsymbol{\psi}_{i_2}^* \boldsymbol{\psi}_{j} \in \mathbb{R}$.
Since the addition of a random variable $X_j$ results in increasing the variance of the metric at $j$, the maximum values in both the numerator and the denominator for the OMP case are likely to be larger than those of the SOMP case.     
Compared with the SOMP case, the relative rate of increase  in the numerator is higher than that of increase in the denominator with  high possibility because $|\boldsymbol{\psi}_{j}|^4$ is larger than $|\boldsymbol{\psi}_{j}^*\boldsymbol{\psi}_{i}|^2$ for $i \neq j$ in general for a random sensing matrix.
 Consequently,   we can infer that the normalized coherence metric of  SOMP is likely to be less than that of  OMP. We validate this inference by simulations for random $\mathbf{\Phi}$, $\cS$, and $\cS_o$ in \sref{subsec:analysis}.      

Although this explains the superiority of  SOMP over  OMP, we can see that 
the probability of $\rho^{\mathrm{(S)}}_{T \rightarrow \infty}(\cS, \cS_o, \mathbf{\Phi})<1$ does not become one in general.
This means that  SOMP can not guarantee the perfect recovery even when an infinite number of snapshots are measured for the recovery. 

Unlike  SOMP,  DSOMP can always guarantee perfect recovery as $T$ becomes larger as shown in the following theorem.

\begin{thm} \label{thm:DSOMP_convergence}
As $T \rightarrow \infty$, $\rho^{\mathrm{(DS)}}(\cS, \cS_o, \mathbf{\Phi}_1,...,\mathbf{\Phi}_T, \bG)$ in \eqref{def_rho_DSOMP0} converges to a deterministic value $\rho^{\mathrm{(DS)}}_{T \rightarrow \infty}(N,M,L,L_o)$ that has an upper bound as
\begin{equation}\label{def_rho_DSOMP_limit}
\begin{split}
\rho^{\mathrm{(DS)}}(\cS, \cS_o, \mathbf{\Phi}_1,...,\mathbf{\Phi}_T, \bG)   
 &\rightarrow \rho^{\mathrm{(DS)}}_{T \rightarrow \infty}(N,M,L,L_o) \\
&\leq  \left( 1 + \frac{\Nrf - \Nt +  (\Nrf - L_o)(\Nt- L_o-1)}{(\Nt - \Nrf) ( L - L_o )} \right)^{-1},  \\
\end{split}
\end{equation}
and this upper bound is always less than or equal to one for any $\Nt \geq \Nrf \geq L > L_o$, i.e,.  perfect recovery is guaranteed.  
\end{thm} 
\begin{IEEEproof}
See Appendix  \ref{app:proof_theorem_2}.
\end{IEEEproof}

In \sref{sec:sim_results}, we show by simulations that a reasonably small $T$ can provide a significant gain over  SOMP even in the extreme case when $\Nrf = L$.  
In addition, the converged value of  DSOMP $\rho^{\mathrm{(DS)}}_{T \rightarrow \infty}(N,M,L,L_o)$  is not a function of sensing matrices $\mathbf{\Phi}_1,...,\mathbf{\Phi}_T$ while that of  SOMP $\rho^{\mathrm{(S)}}_{T \rightarrow \infty}(\cS, \cS_o, \mathbf{\Phi})$ depends on the sensing matrix $\mathbf{\Phi}$. 
In other words, in contrast to the SOMP case where  sophisticated sensing matrix  design is crucial to the estimation performance,  DSOMP can reduce the necessity of the elaborate design of the sensing matrix as $T$ increases. 

Similar to the DSOMP case, 
 DCOMP  can also  guarantee the perfect recovery. The perfect recovery condition of the DCOMP case is shown in the following theorem.

\begin{thm}\label{thm:rho_def_DCOMP}
Let $\cS$ be an optimal support set with $|\cS| = L$ and $\bG$ be an ideal channel gain matrix.
Suppose that $\cS_o \subset \cS $ with $|\cS_o| = L_o$ was perfectly recovered  at the previous $L_o$ iterations  in the DCOMP algorithm.  
In the noiseless case, one of the elements in the optimal support $\cS$ can be perfectly recovered at the $(L_o+1)$-th iteration if and only if
\begin{equation}\label{def_rho_DCOMP}
\begin{split}
\rho ^{\mathrm{(DC)}} (\cS, \cS_o, \mathbf{\Phi}_{1},...,\mathbf{\Phi}_{T},\bG)  &=  \frac{ \max_{j \in \cN \setminus \cS}   \sum_{t=1}^{T} \boldsymbol{\phi}_{t,j}^*   \bQ_{t,\cS,\cS_o}^{\mathrm{(DC)}}  \boldsymbol{\phi}_{t,j}}   {\max_{j \in \cS \setminus \cS_o}  \sum_{t=1}^{T} \boldsymbol{\phi}_{t,j}^*   \bQ_{t,\cS,\cS_o}^{\mathrm{(DC)}}  \boldsymbol{\phi}_{t,j} } < 1,
\end{split}
\end{equation} 
where $\bQ_{t,\cS,\cS_o}^{\mathrm{(DC)}} =
    \mathbf{\Phi}_{t,\cS} \bg_{t,\cS} \bg_{t,\cS}^* \mathbf{\Phi}_{t,\cS}^*  - \bP_{t,\cS_o} \mathbf{\Phi}_{t,\cS} \bg_{t,\cS} \bg_{t,\cS}^* \mathbf{\Phi}_{t,\cS}^* \bP_{t,\cS_o}$ and  $ \bP_{t,\cS_o} =  \mathbf{\Phi}_{t,\cS_o} \mathbf{\Phi}_{t,\cS_o}^{\dagger}$.
\end{thm}

\begin{IEEEproof}
From the definition of the residual matrix $\bV_t$ in \algoref{alg:DCOMP}, the selection rule of  DCOMP can be represented as 
\begin{equation}\label{DCOMP_sel_criterion}
\begin{split}
  j^{\mathrm{(opt)}} = \arg \max_{j\in \cN} \sum_{t=1}^{T}  \boldsymbol{\phi}_{t,j}^* \bV_{t}  \boldsymbol{\phi}_{t,j} 
&= \arg \max_{j \in \cN} \sum_{t=1}^{T} \boldsymbol{\phi}_{t,j}^*   \bQ_{t,\cS,\cS_o}^{\mathrm{(DC)}}  \boldsymbol{\phi}_{t,j},
\end{split}
\end{equation}
and this completes the proof.
\end{IEEEproof}
Let $\bQ_{t,\cS,\cS_o}^{\mathrm{(DS)}} =
    \left( \bI - \bP_{t,\cS_o} \right)   \mathbf{\Phi}_{t,\cS} \bg_{t,\cS} \bg_{t,\cS}^* \mathbf{\Phi}_{t,\cS}^* \left( \bI - \bP_{t,\cS_o} \right)$. 
Notice that,  if $\bQ_{t,\cS,\cS_o}^{\mathrm{(DC)}}$ is replaced by $\bQ_{t,\cS,\cS_o}^{\mathrm{(DS)}} $,
then \thmref{thm:rho_def_DCOMP} becomes identical to \thmref{thm:rho_def_DSOMP}.
It is also worthwhile to note that  both  DSOMP and  DCOMP select the same support at the first iteration because $\bQ_{t,\cS,\cS_o}^{\mathrm{(DS)}} = \bQ_{t,\cS,\cS_o}^{\mathrm{(DC)}} $ for $\cS_o=\emptyset$. The superiority of  DCOMP over  DSOMP starts from the second iteration, which can be inferred from the following theorem.  

\begin{thm} \label{thm:DCOMP_convergence}
$\rho^{\mathrm{(DC)}}(\cS, \cS_o, \mathbf{\Phi}_1,...,\mathbf{\Phi}_T,\bG)  $ in \eqref{def_rho_DCOMP} converges to a deterministic value as $T \rightarrow \infty$, and the deterministic value $\rho^{\mathrm{(DC)}}_{T \rightarrow \infty}(\Nt, \Nrf, L, L_o)  $  always satisfies
\begin{equation}\label{DCOMPvsDSOMP}
 \rho^{\mathrm{(DC)}}_{T \rightarrow \infty}(\Nt, \Nrf, L,L_o)   \leq  \rho^{\mathrm{(DS)}}_{T \rightarrow \infty}(\Nt, \Nrf, L, L_o), 
\end{equation}
for any $\Nt \geq \Nrf \geq L > L_o$.
\end{thm}  

\begin{IEEEproof}
See Appendix  \ref{app:proof_theorem_4}.
\end{IEEEproof}

Until now, we considered a time-varying analog combining matrix and its associated digital combining matrix. One question arises: if the analog combiner is fixed and only the digital combiner is time-varying, does the time variance of the baseband combiner improve the recovery performance? The following proposition shows that  the time-varying baseband combining matrix itself does not improve the performance when the analog combining matrix is fixed. 

\begin{prop}\label{prop:time_varying_Wbb}
Let $\mathbf{\bW}_{\mathrm{BB},t} = \bU_t \left( \mathbf{\bW}_{\mathrm{RF}} \mathbf{\bW}_{\mathrm{RF}}^* \right)^{-\frac{1}{2}}$ for a fixed $\mathbf{\bW}_{\mathrm{RF}}$ and a random unitary matrix $\bU_t \in \mathbb{C}^{M \times M}$  such that  $\mathbf{\Phi}_t = \mathbf{\bW}_{\mathrm{BB},t} \mathbf{\bW}_{\mathrm{RF}} \bA$  constitutes a tight frame for any $t$. 
Then, $\rho^{\mathrm{(DS)}}(\cS,\cS_o,\mathbf{\Phi}_{1},...,\mathbf{\Phi}_{T},\bG)  $ of  DSOMP becomes identical to $\rho^{\mathrm{(DS)}}(\cS,\cS_o,\mathbf{\Phi}_{0},...,\mathbf{\Phi}_{0},\bG)  $ of  SOMP  where $\mathbf{\Phi}_{0} =   \left( \mathbf{\bW}_{\mathrm{RF}} \mathbf{\bW}_{\mathrm{RF}}^* \right)^{-\frac{1}{2}} \mathbf{\bW}_{\mathrm{RF}} \bA$. 
\end{prop} 
\begin{IEEEproof}
Each term in both the numerator and the denominator in \eqref{def_rho_DSOMP_re} becomes
\begin{equation}\label{proof_varyingBB}
\begin{split}
\boldsymbol{\phi}_{t,j}^*   \bQ_{t,\cS,\cS_o}^{\mathrm{(DS)}}  \boldsymbol{\phi}_{t,j} 
& =  \boldsymbol{\phi}_{t,j}^* \left( \bI - \bP_{t,\cS_o} \right)   \mathbf{\Phi}_{t,\cS}  \bg_{t,\cS}  \bg_{t,\cS}^* \mathbf{\Phi}_{t,\cS }^* \left( \bI - \bP_{t,\cS_o} \right) \boldsymbol{\phi}_{t,j} \\
&= \boldsymbol{\phi}_{0,j}^* \left( \bI - \bP_{0,\cS_o} \right)   \mathbf{\Phi}_{0,\cS}  \bg_{t,\cS}  \bg_{t,\cS}^* \mathbf{\Phi}_{0,\cS }^* \left( \bI - \bP_{0,\cS_o} \right) \boldsymbol{\phi}_{0,j},
\end{split} 
\end{equation}
because $\mathbf{\Phi}_t = \bU_t \mathbf{\Phi}_0 $ and $\bP_{t,\cS_o} = \bU_t \bP_{0,\cS_o} \bU_t^*$ for all $t$. 
Since this is the same as in the SOMP case where $\mathbf{\Phi}_1 = \cdots = \mathbf{\Phi}_T  = \mathbf{\Phi}_0 $,   the proof is completed. 
\end{IEEEproof}
The result in \propref{prop:time_varying_Wbb} is  beneficial especially to the wideband case where the analog combining matrix must be fixed over frequency. 
According to \propref{prop:time_varying_Wbb}, we will use the frequency-invariant baseband combining matrix when modifying the algorithm for the wideband case. This will be discussed in more detail in   \sref{subsec:extension_WB}.

\section{Extension to other scenarios} \label{sec:extension_to_others}
In \sref{sec:two_main_ideas}, we  considered a narrowband system and assumed that an MS has a single antenna. In this section, we first extend the work to the multiple-MS-antenna case and then to the wideband case. 

\subsection{Extension to the multiple-MS-antenna case}\label{subsec:extension_multipleMSant}

\begin{figure}[!t]
	\centerline{\resizebox{0.6\columnwidth}{!}{\includegraphics{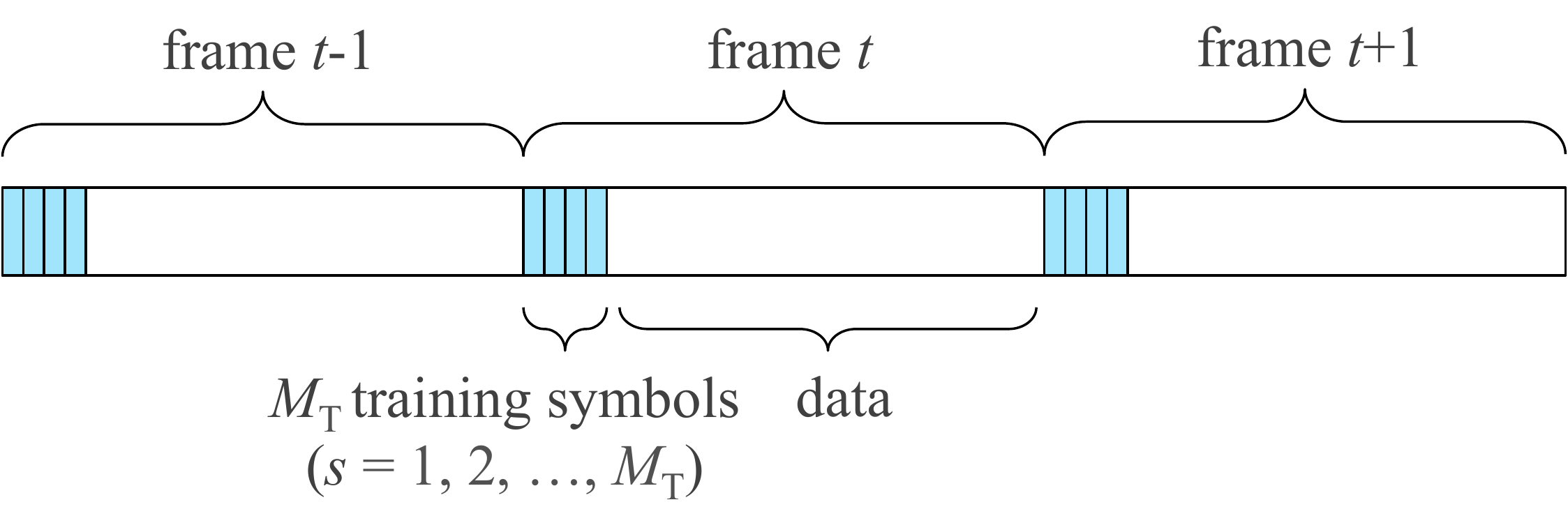}}}
	\caption{Frame structure in the case when an MS has hybrid structure with multiple RF chains and antenna. Consecutive $M_{\mathrm{T}}$ training symbols are inserted in every frame, and each training symbol is transmitted one at a time via its dedicated RF chain.}
	 \label{fig:frame_structure_multipleMSant}
\end{figure}

In this subsection, we extend the work in \sref{sec:two_main_ideas} to the case where the MS has hybrid architecture with multiple antennas and RF chains. The signal  model is shown in \figref{fig:frame_structure_multipleMSant} where $\NantT$, $\NrfT$, $\NantR$, and $\NrfR$ denote the number of transmit antennas, transmit RF chains, receive antennas, and receive RF chains. 
We assume that consecutive $\NrfT$ training symbols are used in each frame as shown in \figref{fig:frame_structure_multipleMSant}, and the symbol index per frame is denoted by $s$. We assume that the duration of $\NrfT$ symbols is less than the channel coherence time, i.e., the channel is invariant during $\NrfT$ consecutive symbol transmission. 
Let $\theta_{\ell}$ be the angle of departure (AoD) of the $\ell$-th path.
The channel model for the single-MS-antenna case in \eqref{eq:ch_model_org} can be extended to the multiple-MS-antenna case as
\begin{equation}\label{eq:ch_model_multipleMSant}
\begin{split}
\bH_{t}  &= \sum_{\ell=1}^{L} g_{t,\ell} \ba_{{\rm{R}}} ( \phi_{\ell})  \ba^{*}_{{\rm{T}}} (\theta_{\ell} ).
\end{split}
\end{equation}
Similarly to \eqref{eq:ch_model_CS}, this can also be  represented in the compressive sensing framework as $\bH_{t}  =\bAdr \bGdt \bAdt^*$,  
where $\bAdr \in \mathbb{C}^{\NantR \times D_{\mathrm{R}}}$ and $\bAdt \in \mathbb{C}^{\NantT \times D_{\mathrm{T}}}$ are dictionary matrices associated with AoA and AoA,  and  $\bGdt \in \mathbb{C}^{D_{\mathrm{R}} \times D_{\mathrm{T}}}$ is a sparse matrix with $L$ nonzero elements with associated with complex channel path gains.

At each symbol $s$ at frame $t$, one training signal is transmitted through one transmit RF chain. Let $\bW_{t,s}$  be a combiner and  $\bf_{t,s}$ a precoder at frame $t$ and symbol $s$. 
Assuming that the training symbol is known to the BS, thereby  omitted as in the single-MS-antenna case, 
the received signal at  baseband is represented as
\begin{equation}\label{eq:rx_bb_signal_multipleMSant}
\begin{split}
\by_{t,s}  = \bW_{t,s} \bAdr \bGdt \bAdt^* \bf_{t,s}  + \bW_{t,s} \bn_{t,s}.
\end{split}
\end{equation}

There are four different types with respect to how to apply  $\bW_{t,s}$ and $\bf_{t,s}$ within a frame:  
1) fixed $\bW_{t,s}$ and fixed $\bf_{t,s}$, 2) time-varying $\bW_{t,s}$ and fixed $\bf_{t,s}$, 3) fixed $\bW_{t,s}$ and time-varying $\bf_{t,s}$, and 4) time-varying $\bW_{t,s}$ and time-varying $\bf_{t,s}$.

In the first approach, all the precoders are the same within a frame, and so are the combiners such that $\bff_{t,1}=\cdots = \bff_{t,\NrfT} = \bff_{t}$ and $\bW_{t,1}=\cdots = \bW_{t,\NrfT} = \bW_{t}$. Since these combiners and precoders are constant within a frame in this approach, averaging the received signals at baseband within a frame can increase the effective signal-to-noise ratio (SNR). 
The averaged received signal is represented as
\begin{equation}\label{eq:avg_received_signal}
\begin{split}
\bar{\by}_{t}   &=\frac{1}{\NrfT} \sum_{s=1}^{\NrfT} \by_{t,s} = \bW_{t} \bAdr \bGdt \bAdt^* \bff_{t}  +\bW_{t}    \bar{\bn}_{t},  
\end{split}
\end{equation}
where $\bar{\bn}_{t}   =\frac{1}{\NrfT} \sum_{s=1}^{\NrfT} \bn_{t,s}   $. Note that $\bar{\bn}_{t}\sim \mathcal{CN}\left(\mathbf{0},\frac{\sigma^2}{\NrfT}\bI \right)$, which means that this process averages out the noise, increasing the effective SNR. 
The signal model per frame in \eqref{eq:avg_received_signal} can be transformed into 
\begin{equation}\label{eq:avg_received_signal_r1}
\begin{split}
\bar{\by}_{t} &= \left( \bff_{t}^T \otimes \bW_{t}  \right  )  \left(  \bAdt^C \otimes \bAdr  \right  ) \mathrm{vec}(\bGdt) +\bW_{t} \bar{\bn}_{t}.
\end{split}
\end{equation}

In the second approach, where only combiners vary and  precoders are fixed within a frame such that $\bff_{t,1}=\cdots = \bff_{t,\NrfT} = \bff_{t}$, the first step is to stack $\by_{t,1},...,\by_{t,\NrfT}$ in rows. This row-wise stack yields a $\NrfR  \NrfT \times 1$ vector per frame, 
\begin{equation}\label{eq:multipleMSant_2nd}
\begin{split}
\tilde{\by}_{t,\mathrm{agg}} 
&= \tilde{\bW}_{t,\mathrm{agg}}  \bAdr \bG_{t} \bAdt^* \bff_{t} + \tilde{\bn}_{t,\mathrm{agg}},
\end{split}
\end{equation}
where $\tilde{\bW}_{t,\mathrm{agg}} = \begin{bmatrix} \bW_{t,1}^T & \cdots & \bW_{t,\NrfT}^T \end{bmatrix}^T$ and $\tilde{\bn}_{t,\mathrm{agg}} =  \begin{bmatrix} \bn_{t,1}^T \bW_{t,1}^T  & \cdots & \bn_{t,\NrfT}^T \bW_{t,\NrfT}^T  \end{bmatrix}^T$.
The signal model in \eqref{eq:multipleMSant_2nd} can be rewritten as 
\begin{equation}\label{eq:multipleMSant_2nd_r1}
\begin{split}
\tilde{\by}_{t,\mathrm{agg}}  
&= \left( \bff_{t}^T \otimes \tilde{\bW}_{t,\mathrm{agg}}    \right  ) \left(  \bAdt^C \otimes \bAdr  \right  ) \mathrm{vec}(\bGdt) + \tilde{\bn}_{t,\mathrm{agg}}.
\end{split}
\end{equation}

The third approach changes only precoders,  and  combiners are fixed within a frame such that  $\bW_{t,1}=\cdots = \bW_{t,\NrfT} = \bW_{t}$. In contrast to the second approach, 
$\by_{t,1},...,\by_{t,\NrfT}$ are stacked in columns. This column-wise stack makes the aggregate signal model per frame as 
\begin{equation}\label{eq:multipleMSant_3rd}
\begin{split}
\tilde{\bY}_{t,\mathrm{agg}} &= \bW_{t} \bAdr \bG_{\mathrm{D},t} \bAdt^* \tilde{\bF}_{t,\mathrm{agg}}  + \tilde{\bN}_{t,\mathrm{agg}},
\end{split}
\end{equation}
where $\tilde{\bY}_{t,\mathrm{agg}} = \begin{bmatrix} \by_{t,1} & \cdots & \by_{t,\NrfT} \end{bmatrix}$, 
$\tilde{\bF}_{t,\mathrm{agg}}  = \begin{bmatrix} \bff_{t,1} & \cdots & \bff_{t,\NrfT} \end{bmatrix}$, and 
$\tilde{\bN}_{t,\mathrm{agg}} = \bW_{t}^*  \begin{bmatrix} \bn_{t,1} & \cdots & \bn_{t,\NrfT} \end{bmatrix}$.
By using vectorization, the matrix $\tilde{\bY}_{t,\mathrm{agg}}$ in \eqref{eq:multipleMSant_3rd} can be transformed into a 
vector form as 
\begin{equation}\label{eq:multipleMSant_3rd_r1}
\begin{split}
\tilde{\by}&_{t,\mathrm{agg}} = \mathrm{vec}(\tilde{\bY}_{t,\mathrm{agg}}) 
 = \left( \tilde{\bF}_{t,\mathrm{agg}} ^T \otimes \bW_{t}  \right  )  \left(  \bAdt^C \otimes \bAdr  \right  ) \mathrm{vec}(\bGdt) +\mathrm{vec}(\tilde{\bN}_{t,\mathrm{agg}}).
\end{split}
\end{equation}

For the last approach, both  precoders and combiners change within a frame. Like the second approach, we stack $\by_{t,1},...,\by_{t,\NrfT}$ in rows. Then, the row-wise stacked vector $\tilde{\by}_{t,\mathrm{agg}}$ becomes
\begin{equation}\label{eq:multipleMSant_4th}
\begin{split}
\tilde{\by}_{t,\mathrm{agg}} 
&= \left( \tilde{\bF}_{t,\mathrm{agg}} \stackrel{(\NrfT)}{\odot}   \tilde{\bW}_{t,\mathrm{agg}}^T  \right  )^T  \left(  \bAdt^C \otimes \bAdr  \right  ) \mathrm{vec}(\bGdt) +\tilde{\bn}_{t,\mathrm{agg}}. 
\end{split}
\end{equation}

Note that the signal models for all four different approaches in \eqref{eq:avg_received_signal_r1}, \eqref{eq:multipleMSant_2nd_r1},  \eqref{eq:multipleMSant_3rd_r1}, and \eqref{eq:multipleMSant_4th} can be generalized as $\tilde{\by}_{t,\mathrm{agg}}  = \mathbf{\Theta}_{t,\mathrm{agg}}  \mathbf{A}_{\mathrm{agg}} \bg_{t,\mathrm{agg}} + \tilde{\bn}_{t_{t,\mathrm{agg}}  }$
where $\bg_{t,\mathrm{agg}} = \mathrm{vec}(\bGdt)$, $ \mathbf{A}_{\mathrm{agg}} = \bAdt^C \otimes \bAdr$, and $
\mathbf{\Theta}_{t,\mathrm{agg}}$ are dependent on each approach.   
This generalized signal model has the same format as that of the single-MS-antenna case in  \eqref{eq:rx_bb_model_vec}. Consequently, the algorithm proposed for the single-MS-antenna case such as DCOMP can be used without any modification in the multiple-MS-antenna case regardless of different approaches.

\subsection{Extension to wideband systems}\label{subsec:extension_WB}

\begin{algorithm}[t]
\caption{Wideband DCOMP (WB-DCOMP)}
\label{alg:WB_DCOMP}
\begin{algorithmic}
\State Input:  $\mathbf{\Phi}_1,..., \mathbf{\Phi}_T, \by_{1,1}, ..., \by_{T,K}, L $
\State Initialize: $ \bV_{t}=\hat{\bR}_{\by,t} = \frac{1}{K} \sum_{k=1}^K \by_{t,k} \by_{t,k}^*,  \mathcal{S}=\O, \hat{\bR}_{\bg} = \mathbf{0}$
\FOR {$n =1:L $}
\State $j = \arg \max_{i} \sum_{t=1}^{T}   \boldsymbol{\phi}_{t,i}^*\bV_t \boldsymbol{\phi}_{t,i} $
\State $\mathcal{S} = \mathcal{S}  \cup \{ j \}$
\State $\bV_t = \bR_{\by,t} - \mathbf{\Phi}_{t,\mathcal{S}}  \mathbf{\Phi}^{\dagger}_{t,\mathcal{S}}  \bR_{\by,t}  \left( \mathbf{\Phi}_{t,\mathcal{S}} \mathbf{\Phi}^{\dagger}_{t,\mathcal{S}}   \right)^{*},    \; \forall t$
\ENDFOR
\State $[\hat{\bR}_{\bg}]_{\mathcal{S},\mathcal{S}} =  \frac{1}{T} \sum_{t=1}^T \mathbf{\Phi}^{\dagger}_{t,\mathcal{S}}  \bR_{\by,t}  \left(\mathbf{\Phi}^{\dagger}_{t,\mathcal{S}}   \right)^{*}$ 
\State Output: $\hat{\bR}_{\bg} $
\end{algorithmic}
\end{algorithm}

In this subsection, we extend our work in the narrowband case into the wideband case. Since the same algorithm can be used for both the single-MS-antenna case and the multiple-MS-antenna case as shown in \sref{subsec:extension_multipleMSant}, we focus on the single antenna case for the sake of exposition. 

Compared to the narrowband channel model in \eqref{eq:ch_model_org}, 
we adopt the delay-$d$ MIMO channel model  
 for the wideband OFDM systems with $K$ subcarriers \cite{SchniterSayeed2014,Alkhateeb2015a}. Let $T_s$ be the sampling period, $\tau_\ell$ be the delay of the $\ell$-th path, $N_{\mathrm{CP}}$ be the cyclic prefix length, and $p(t)$ denote a filter comprising the combined effects of pulse shaping and other analog filtering. 
The delay-$d$ MIMO channel matrix is modeled as  
\begin{equation}\label{WB_ch_model_CIR}
\bh_{t}[d]  = \sum_{\ell=1}^{L} g_{\ell,t} p(dT_s - \tau_{\ell}) \ba ( \phi_{\ell}) \;\; \text{for} \;\; d=0,..., N_{\mathrm{CP}-1}, 
\end{equation}
and the channel frequency response matrix at each subcarrier $k$ can be expressed as
\begin{equation}\label{WB_ch_model_CFR}
\begin{split}
\bh_{t,k}  
&=\sum_{\ell=1}^{L} g_{\ell,t} c_{\ell,k} \ba( \phi_{\ell})  ,
\end{split}
\end{equation}
where $c_{\ell,k} = \sum_{d=0}^{N_{\mathrm{CP}}-1} p(dT_s - \tau_{\ell}) e^{ -\frac{j 2 \pi k d}{K}}$.
This can be  cast in the compressive sensing framework  as
 $\bh_{t,k}  = \bA  \left(\bg_t  \circledcirc \bc_k \right)$, 
where $\bg_{t}$ and $\bc_{k}$ are sparse  vectors with  $L$ nonzero elements associated with $g_{\ell,t}$ and $c_{\ell,k}$.
Note that $\bg_{t}$ and $\bc_{k}$ share the same support for all $t$ and $k$.

An analog combiner $\bW_{\mathrm{RF},t}$ at frame $t$ must be common for all subcarrier $k$.  Since \propref{prop:time_varying_Wbb} shows that  
changing baseband combiners does not  impact the performance as long as the analog combiner is fixed, we  use a common sensing matrix  $\bW_t = \bW_{\mathrm{BB},k} \bW_{\mathrm{RF},t} $ for all subcarriers at frame $t$. 
Then, the signal part of the received signal in the baseband at frame $t$ and subcarrier $k$ is given by 
$\br_{t,k}  =  \bW_{t}  \bA  \left(\bg_{t} \circledcirc \bc_{k} \right)$.

The problem of finding $\bg_{t} \circledcirc \bc_{k}$ is similar to the narrowband case in \sref{sec:two_main_ideas} because $\bg_{t} \circledcirc \bc_{k}$ share the same support for all $t$ and $k$. Therefore, the DCOMP algorithm in \sref{sec:two_main_ideas}  can be directly applied to this problem. This fails, however, to exploit a useful property of sparse wideband channels.
At a fixed subcarrier $k=k_0$, the time domain average of $\br_{t,k_0} \br_{t,k_0}^*$ becomes
\begin{equation}\label{r_cov_time}
\begin{split}
\sum_{t=1}^T \br_{t,k_0}  \br_{t,k_0} ^* &= \sum_{t=1}^T \bW_{t} \bA   \left( \bg_{t} \circledcirc \bc_{k_0} \right)  \left( \bg_{t} \circledcirc \bc_{k_0} \right)^* \bA^* \bW_{t}^*.   
\end{split}
\end{equation} 
Since \eqref{r_cov_time} has a similar form to \eqref{eq:Relation_btw_Ry_Rg_dynamic} of the narrowband case,  DCOMP can be used for this time domain operation. 
The frequency domain average of $\br_{t_0,k} \br_{t_0,k}^*$ at a fixed frame $t=t_0$ can be expressed differently  from  \eqref{r_cov_time} as 
\begin{equation}\label{r_cov_freq}
\begin{split}
\sum_{k=1}^K \br_{t_0,k}  \br_{t_0,k} ^* &= \bW_{t_0} \bA \left( \sum_{k=1}^K   \left( \bg_{t_0} \circledcirc \bc_{k} \right)  \left( \bg_{t_0} \circledcirc \bc_{k} \right)^* \right) \bA^* \bW_{t_0}^* ,
\end{split}
\end{equation} 
which is similar to \eqref{eq:Relation_btw_Ry_Rg} of the narrowband case. Since  COMP is developed based on the signal model in \eqref{eq:Relation_btw_Ry_Rg},  COMP can be used for the frequency domain operation in the wideband case. The different features between the time  and  frequency domain motivate us to apply different approaches to each domain, i.e., COMP to the frequency domain and  DCOMP to the time domain. This combination of two algorithms, which we call WB-DCOMP, is described in \algoref{alg:WB_DCOMP}.  

It is worthwhile to compare  WB-DCOMP to the direct extension of COMP and DCOMP. 
If the sensing matrix $\mathbf{\Phi}$ is varying over both time and frequency, the direct extension of DCOMP provides  the best performance. However, if $\mathbf{\Phi}$ is varying only in the time domain and fixed over frequency,  WB-DCOMP, which is the combination of DCOMP and COMP, can outperform the direct extension of each algorithm in the wideband case. This is demonstrated in \sref{sec:sim_results}.

\begin{figure}[t]
	\centering
	\subfigure[center][{CDF of $\rho^{\mathrm{(S)}}(\cS, \cS_0, \mathbf{\Phi}, \bG)$ when $L_o=0$}]{
		\includegraphics[width=.41\columnwidth]{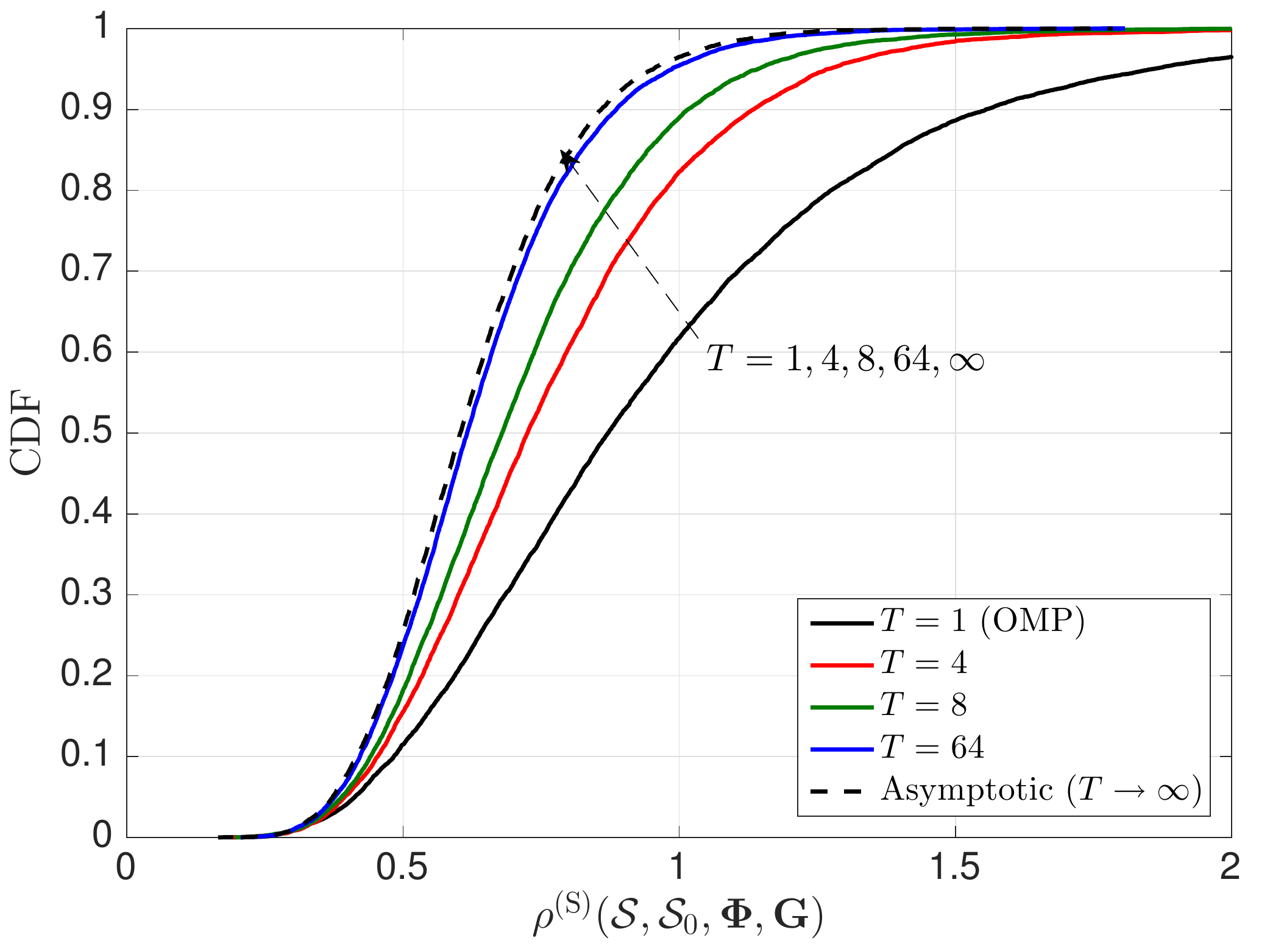}
		\label{fig:Fig_N64_M8_L8_Lo0_SOMP_Tsweep1_4_8_64}}
	\subfigure[center][{CDF of $\rho^{\mathrm{(S)}}(\cS, \cS_0, \mathbf{\Phi}, \bG)$  when $T=1$ or $T=\infty$}]{
		\includegraphics[width=.41\columnwidth]{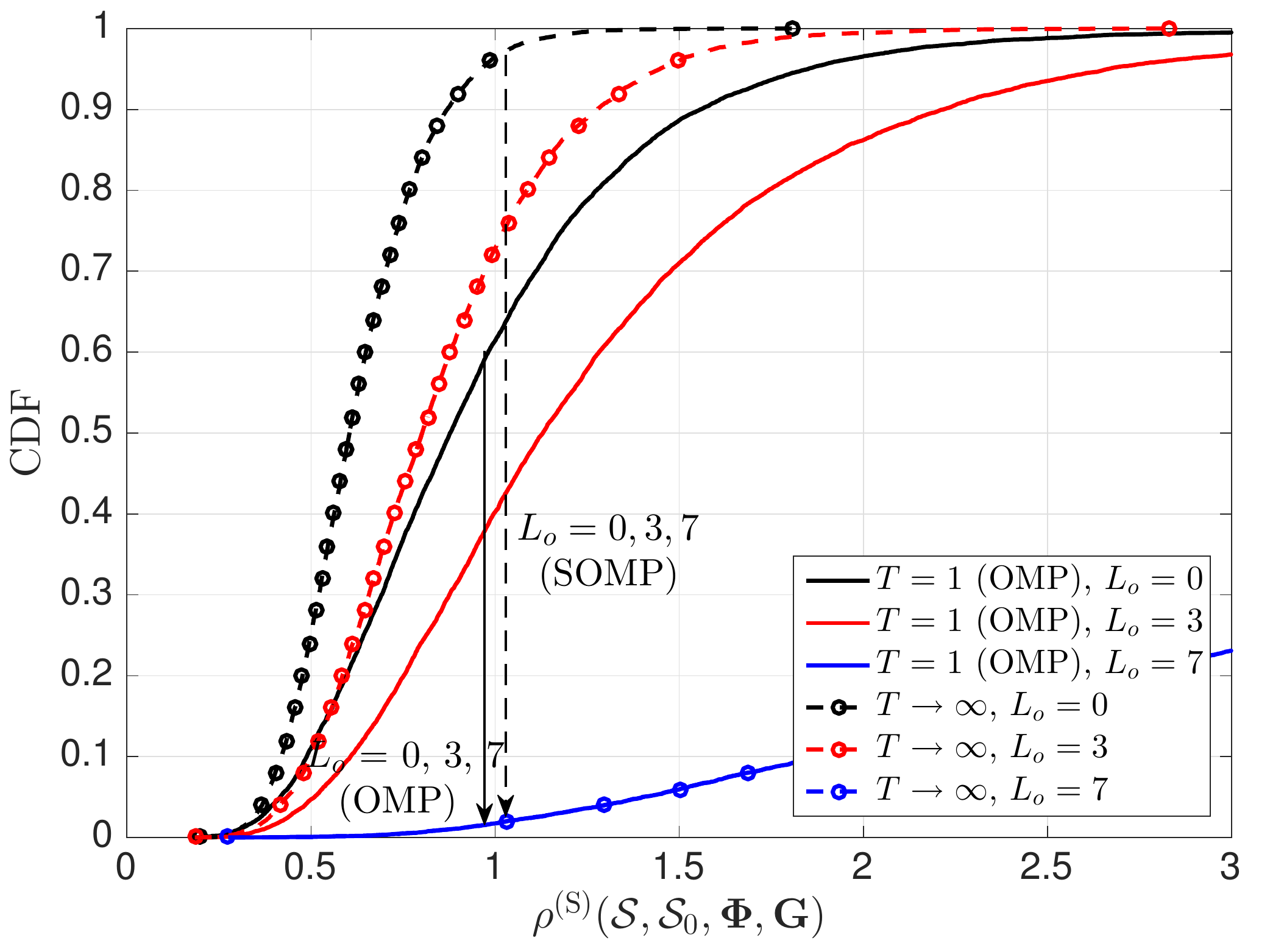}
		\label{fig:Fig_N64_M8_L8_SOMP_L_o_sweep0_3_7}}
	\caption{CDFs of $\rho^{\mathrm{(S)}}(\cS, \cS_0, \mathbf{\Phi},  \bG)$. The optimal support set $\cS$ has $L=8$ elements, the optimally preselected subset $\cS_o$ has $L_o$ elements, $\mathbf{\Phi} \in \mathbb{C}^{M \times N} $ is a fixed sensing matrix of $\Nrf =8$ and $\Nt=64$, and $\bG \in \mathbb{C}^{L \times T}$ is a random matrix whose elements have zero-mean and unit-variance.}
	\label{fig:Fig_N64_M8_L8_SOMP}
\end{figure}

\section{Simulation results}\label{sec:sim_results}

In this section, we  validate our analysis in \sref{sec:theoretical_analysis} in terms of the  recovery success probability in the noiseless case especially when $\Nrf$ has a similar value to $L$. We also present simulation results to demonstrate the performance of the proposed  spatial channel covariance estimation algorithms  for hybrid architecture.

\subsection{Theoretical anaylsis  in the noiseless case}\label{subsec:analysis}

We compare OMP and SOMP in terms of the cumulative distribution function (CDF) of the newly defined metric $\rho^{\mathrm{(S)}}(\cS, \cS_0, \mathbf{\Phi}, \bG)$ in \eqref{def_rho_SOMP}.
\figref{fig:Fig_N64_M8_L8_SOMP} shows   how the CDF of $\rho^{\mathrm{(S)}}(\cS, \cS_0, \mathbf{\Phi}, \bG)$ changes according to $T$ and $L_o$. In \figref{fig:Fig_N64_M8_L8_Lo0_SOMP_Tsweep1_4_8_64}, $\mathrm{Pr}(\rho^{\mathrm{(S)}}(\cS, \cS_0, \mathbf{\Phi}, \bG) <1)$ becomes higher as $T$ gets larger. In addition, as shown in \propref{prop:SOMP_rho}, the CDF converges to that of $\rho^{\mathrm{(S)}}_{T \rightarrow \infty}(\cS, \cS_0, \mathbf{\Phi})$ in \eqref{def_rho_SOMP} regardless of $\bG$. Note that $T=1$ indicates the OMP case.  \figref{fig:Fig_N64_M8_L8_SOMP_L_o_sweep0_3_7} shows the CDF versus $L_o$ in  the OMP case  ($T=1 $) and the asymptotic SOMP case ($T=\infty$). As $L_o$ increases, the gap between  OMP and  SOMP reduces and becomes  zero if $M=L$. In this extreme case where $M=L$, both  OMP and  SOMP do not work properly because  $\mathrm{Pr}(\rho^{\mathrm{(S)}}(\cS, \cS_0, \mathbf{\Phi}, \bG) <1)$ becomes  low in the last iteration.

\begin{figure}[t]
	\centering
	\subfigure[center][{$L_o=0$}]{
		\includegraphics[width=.41\columnwidth]{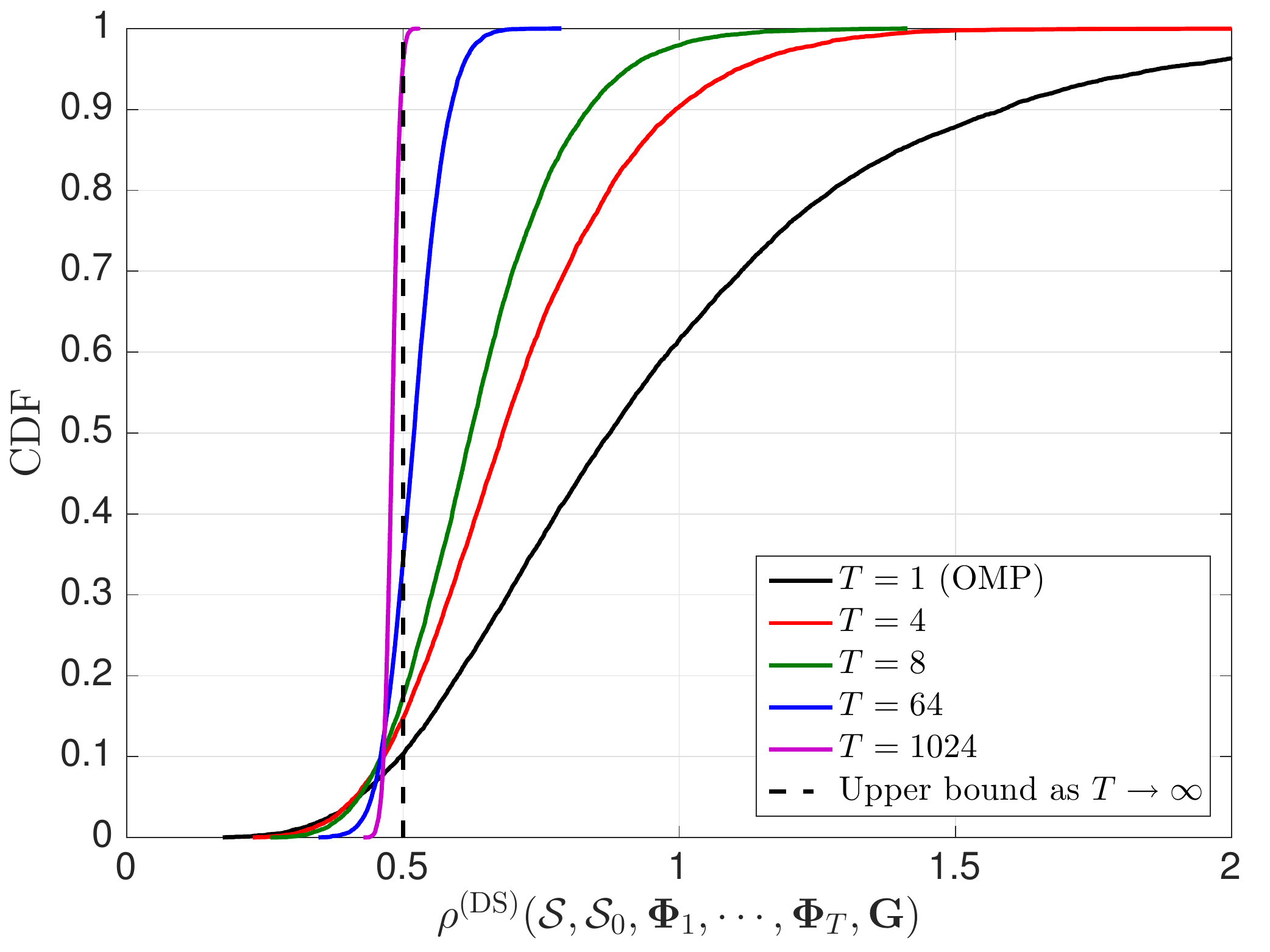}
		\label{fig:Fig_N64_M8_L8_Lo0_DSOMP_Tsweep1_4_8_64_1024}}
	\subfigure[center][{$L_o=7$}]{
		\includegraphics[width=.41\columnwidth]{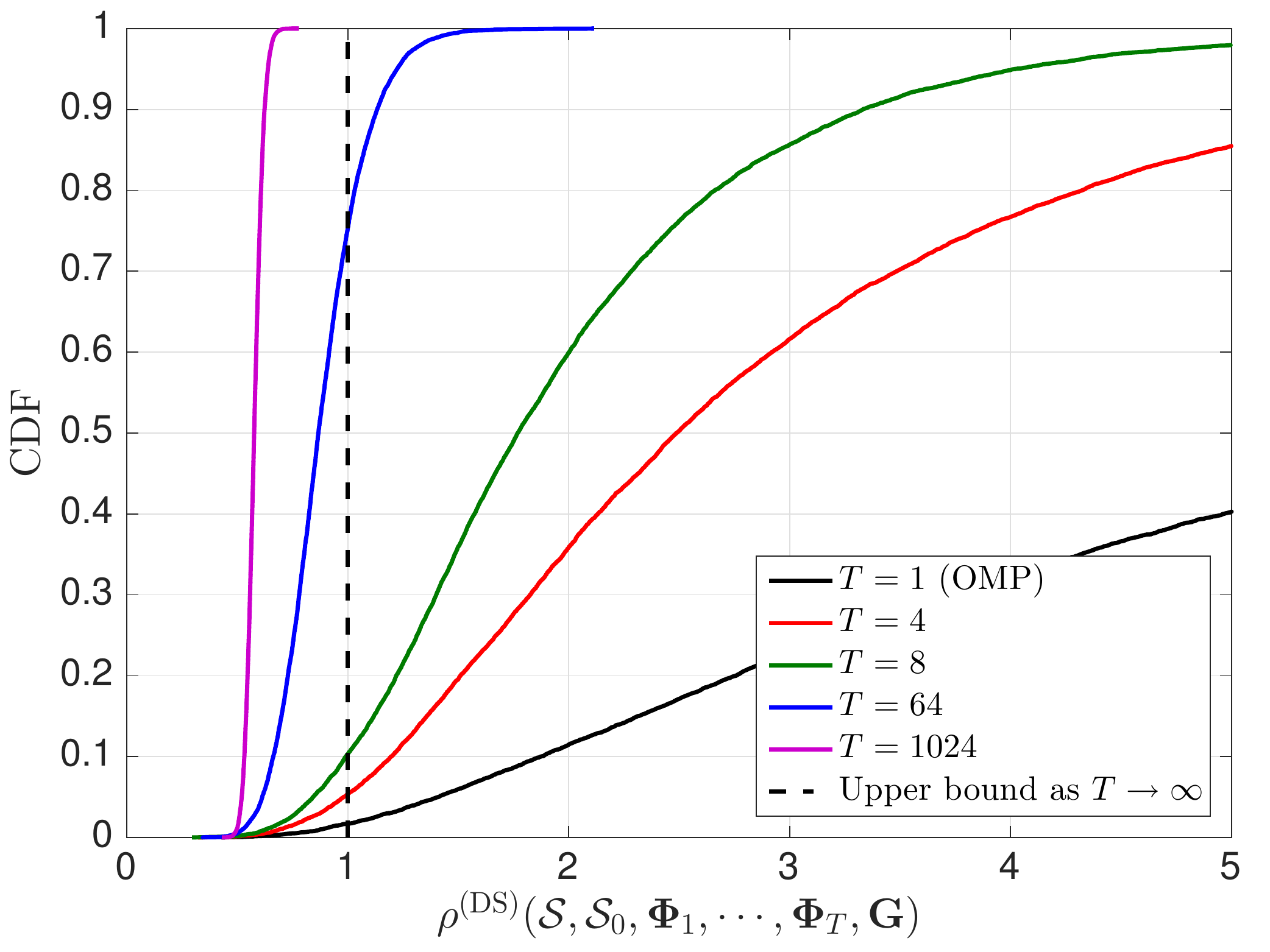}
		\label{fig:Fig_N64_M8_L8_Lo7_DSOMP_Tsweep1_4_8_64_1024}}
	\caption{
	CDFs of $\rho^{\mathrm{(DS)}}(\cS, \cS_0, \mathbf{\Phi}_1,...,\mathbf{\Phi}_T, \bG)$. The optimal support set $\cS$ has $L=8$ elements, the optimally preselected subset $\cS_o$ has $L_o$ elements, $\mathbf{\Phi}_1, ..., \mathbf{\Phi}_T \in \mathbb{C}^{M \times N} $ are time-varying sensing matrices of $\Nrf =8$ and $\Nt=64$, and $\bG \in \mathbb{C}^{L \times T}$ is a random matrix whose elements have zero-mean and unit-variance. The upper bound of $\rho^{\mathrm{(DS)}}(\cS, \cS_0, \mathbf{\Phi}_1,...,\mathbf{\Phi}_T, \bG)$ as $T \rightarrow \infty$ denotes $\rho^{\mathrm{(DS)}}_{T \rightarrow \infty}(N,M,L,L_o)$ in \thmref{thm:DSOMP_convergence}.
	}
	\label{fig:Fig_N64_M8_L8_DSOMP}
\end{figure}

\figref{fig:Fig_N64_M8_L8_DSOMP} shows the CDF of $\rho^{\mathrm{(DS)}}(\cS, \cS_0, \mathbf{\Phi}_1,...,\mathbf{\Phi}_T, \bG)$ in the DSOMP case. 
Unlike the SOMP case, 
$\rho^{\mathrm{(DS)}}(\cS, \cS_0, \mathbf{\Phi}_1,...,\mathbf{\Phi}_T, \bG)$  converges to a deterministic value $\rho^{\mathrm{(DS)}}_{T \rightarrow \infty}(N,M,L,L_o)$ in \eqref{def_rho_DSOMP_limit} for any sensing matrices $\mathbf{\Phi}_1,...,\mathbf{\Phi}_T$ as $T$ increases.  The simulation results in \figref{fig:Fig_N64_M8_L8_DSOMP} also indicate that $\rho^{\mathrm{(DS)}}_{T \rightarrow \infty}(N,M,L,L_o)$ has an  upper bound as shown in  \eqref{def_rho_DSOMP_limit}, which is consistent with the analytical result in    \thmref{thm:DSOMP_convergence}. 
Although the convergence rate becomes slower as $L_o$ increases, $\mathrm{Pr}(\rho^{\mathrm{(DS)}}(\cS, \cS_0, \mathbf{\Phi}_1,..., \mathbf{\Phi}_T, \bG) <1) = 1$ is guaranteed as $T \rightarrow \infty$. Simulation results in \figref{fig:Fig_N64_M8_L8_DSOMP} show that a reasonably small value of $T \ll \infty$ can ensure a perfect recovery.

\begin{figure}[!t]
	\centerline{\resizebox{0.41\columnwidth}{!}{\includegraphics{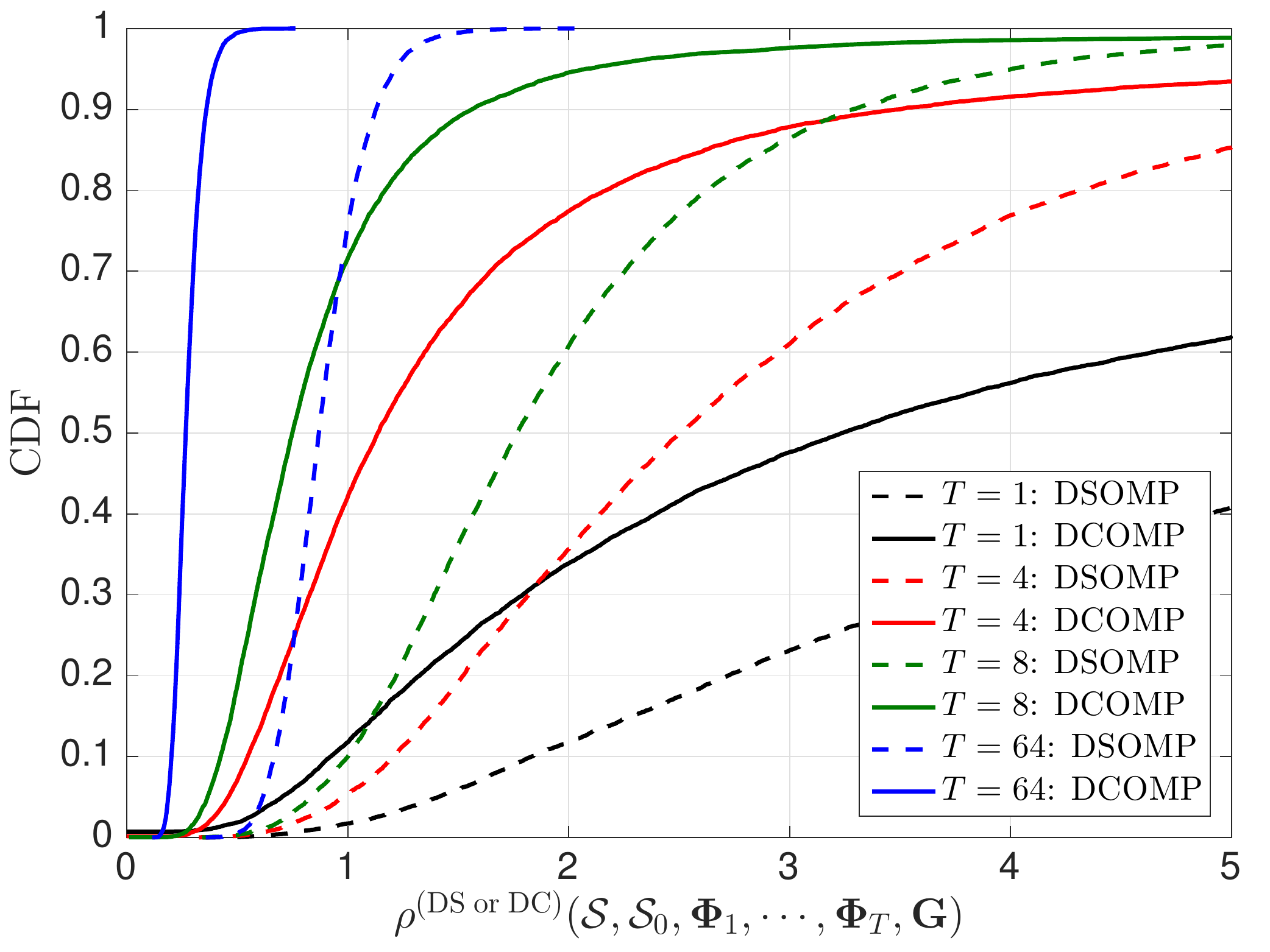}}}
	\caption{Comparison between $\rho^{\mathrm{(DS)}}(\cS, \cS_0, \mathbf{\Phi}_1,...,\mathbf{\Phi}_T, \bG)$ of DSOMP and $\rho^{\mathrm{(DC)}}(\cS, \cS_0, \mathbf{\Phi}_1,...,\mathbf{\Phi}_T, \bG)$ of DCOMP when $\Nt=64, \Nrf =8, L=8$, and $L_o=7$.  }
	\label{fig:Fig_N64_M8_L8_Lo7_DSOMPandDCOMP_Tsweep1_4_8_64}
\end{figure}

\figref{fig:Fig_N64_M8_L8_Lo7_DSOMPandDCOMP_Tsweep1_4_8_64} compares  DSOMP and  DCOMP. As discussed in \sref{sec:theoretical_analysis}, both select the same support at the first iteration, and the superiority of  DCOMP over  DSOMP starts from the second iteration. In \figref{fig:Fig_N64_M8_L8_Lo7_DSOMPandDCOMP_Tsweep1_4_8_64}, the CDF of  $\rho^{\mathrm{(DS)}}(\cS, \cS_0, \mathbf{\Phi}_1,...,\mathbf{\Phi}_T, \bG)$ and $\rho^{\mathrm{(DC)}}(\cS, \cS_0, \mathbf{\Phi}_1,...,\mathbf{\Phi}_T, \bG)$ are compared in the last iteration, i.e., $L_o=7$. 
The figure shows that
the CDF of  the DCOMP case is located on the left side of that of the DSOMP case, which means  DCOMP has a higher success  probability than  DSOMP.

\begin{figure}[t]
	\centering
	\subfigure[center][{DSOMP}]{
		\includegraphics[width=.41\columnwidth]{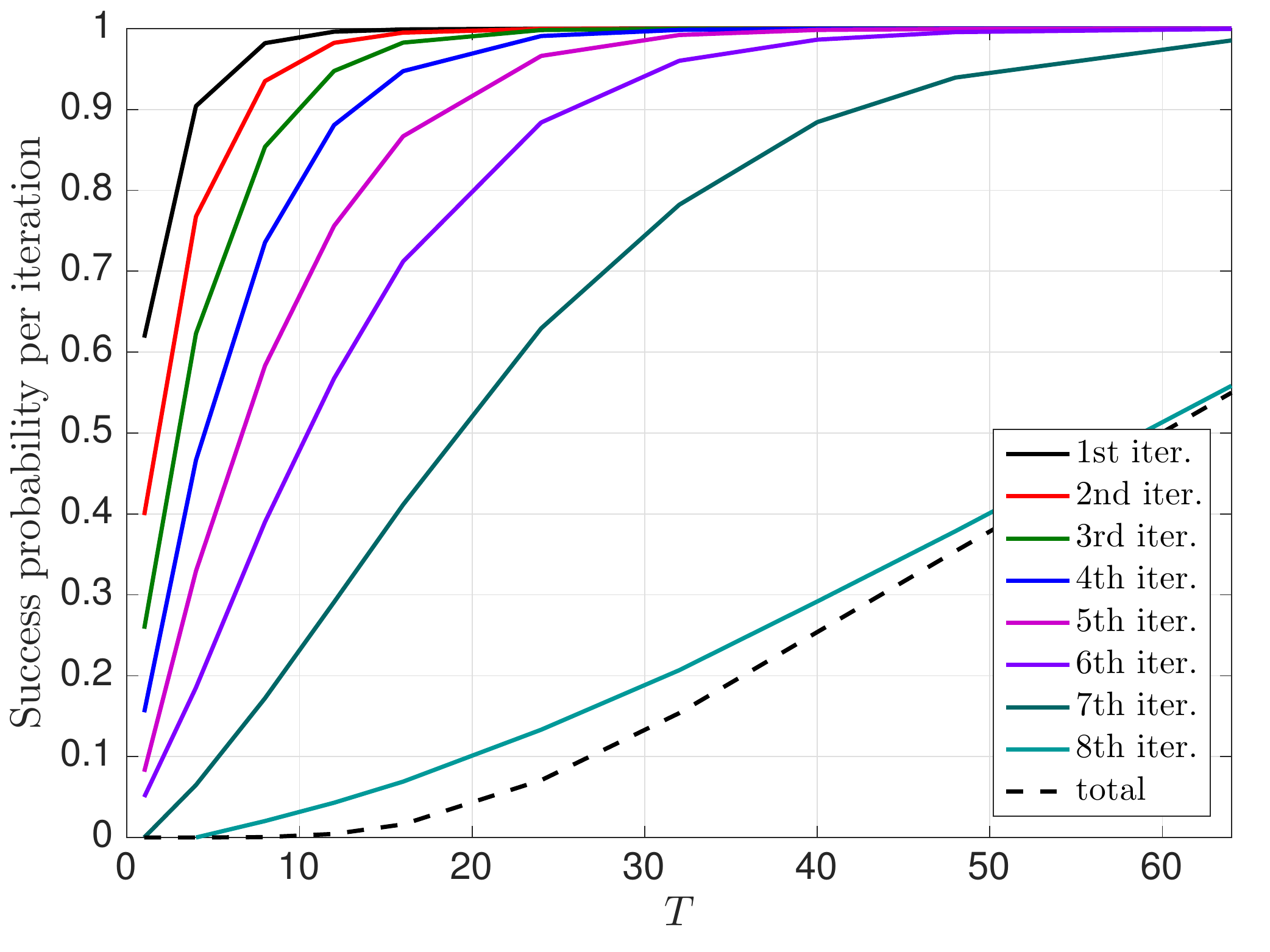}
		\label{fig:Fig_N64_M8_L8_SuccessProb_DSOMP}}
	\subfigure[center][{DCOMP}]{
		\includegraphics[width=.41\columnwidth]{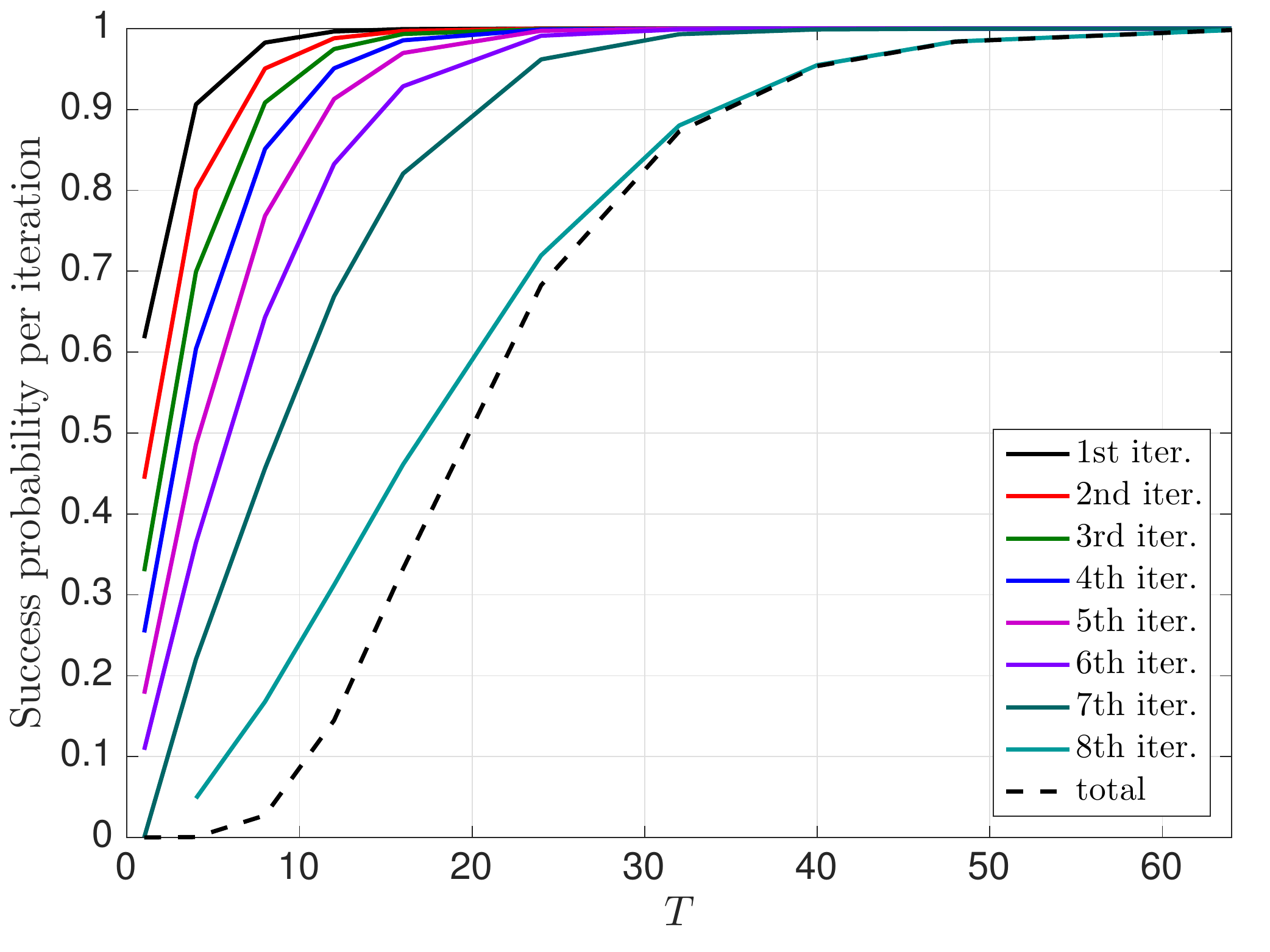}
		\label{fig:Fig_N64_M8_L8_SuccessProb_DCOMP}}
	\caption{Success probability of recovering one of elements in the optimal support set at each iteration for $L_o=1,...,L$ when $\Nt=64, \Nrf =8$,  and  $L=8$.}
	\label{fig:Fig_N64_M8_L8_SuccessProb}
\end{figure}

Simulation results with respect to the success probability per iteration is shown in \figref{fig:Fig_N64_M8_L8_SuccessProb}. As expected, the success probability at the first iteration is identical for both  DSOMP and  DCOMP, but the success probability of  DCOMP becomes higher than that of  DSOMP from the second iteration, leading to the higher success probability in total.  

\subsection{Performance evaluation on the covariance estimation for hybrid architecture} \label{subsec:performance}

In this subsection, we evaluate the performance of the spatial channel covariance estimation. 
The performance metric 
is defined as $\eta = \frac{\mathrm{Tr} \left( \bU^*_{\hat{\bR}_{\bh}} {\bR}_{\bh} \bU_{\hat{\bR}_{\bh}}  \right) } {\mathrm{Tr} \left( \bU^*_{{\bR}_{\bh}} {\bR}_{\bh} \bU_{{\bR}_{\bh}}  \right) }$ \cite{DBLP:journals/corr/HaghighatshoarC15::9}  
where $\bU_{{\bR}_{\bh}}$ and $\bU_{\hat{\bR}_{\bh}}$ are the matrices whose columns are the eigenvectors of the ideal covariance ${\bR}_{\bh}$ and the estimated covariance $\hat{\bR}_{\bh}$. Note that $0 \leq \eta \leq 1$ and a larger $\eta$ indicates a more accurate estimation. 

\begin{figure}[t]
	\centering
	\subfigure[center][$\Nrf = 16$]{
		\includegraphics[width=.41\columnwidth]{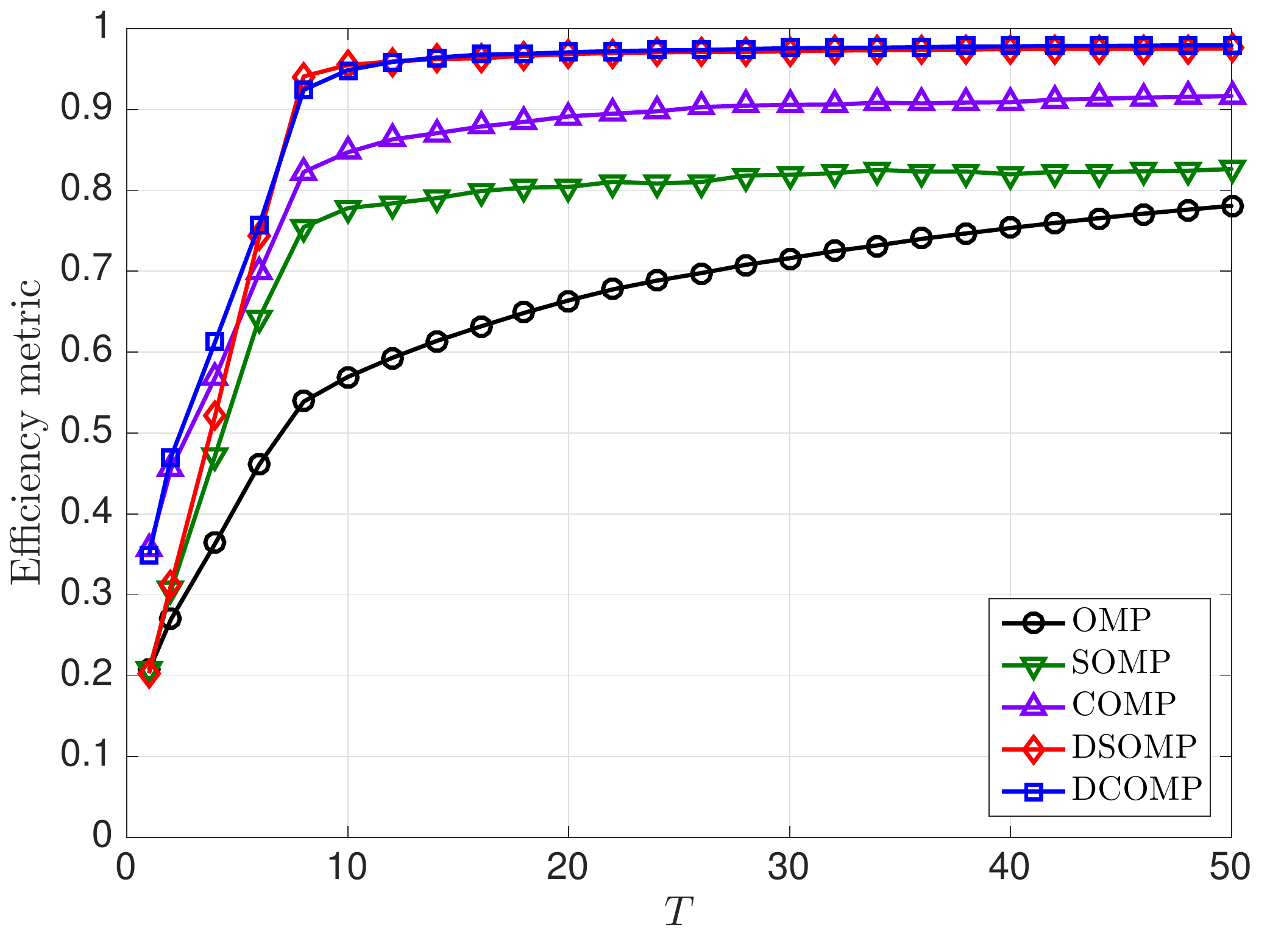}
		\label{fig:NB_SingleAnt_64antennas_16RFchains}}
	\subfigure[center][$\Nrf = 8$]{
		\includegraphics[width=.41\columnwidth]{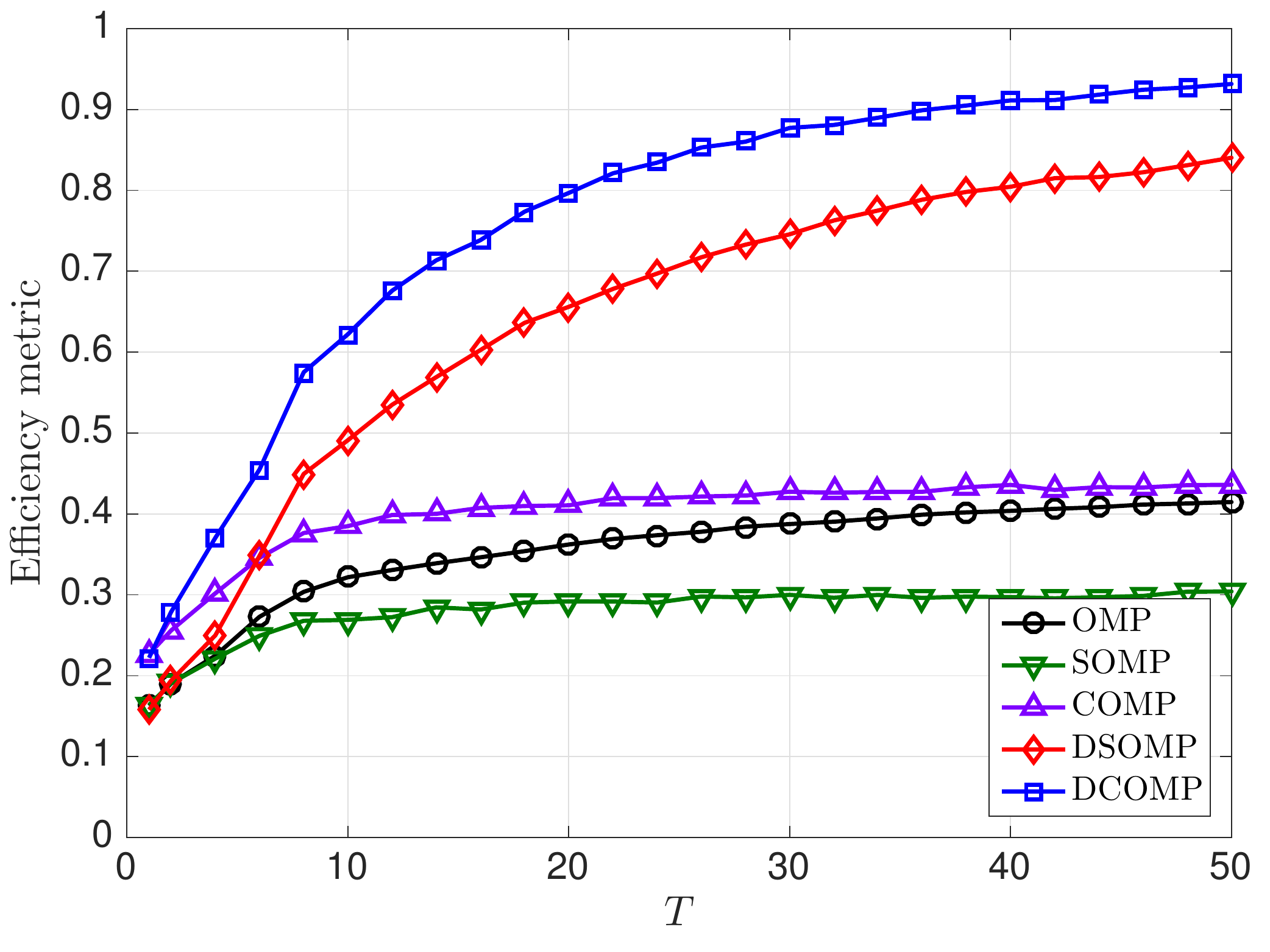}
		\label{fig:NB_SingleAnt_64antennas_8RFchains}}
	\caption{Comparison  among OMP, SOMP, COMP, DSOMP, and DCOMP  when $\Nt = 64$, $D=256$, $L=8$ for the single-MS-antenna case in narrowband systems.}
	\label{fig:NB_SingleAnt_64antennas}
\end{figure}

\figref{fig:NB_SingleAnt_64antennas} shows  simulation results when $\Nt=64, D=256,L=8$, and $\rm{SNR}= 10$ dB for the single-MS-antenna case in narrowband systems. In the fixed combining matrix case in \figref{fig:NB_SingleAnt_64antennas_16RFchains} where $\Nrf = 16$, we can see that the proposed COMP outperforms OMP and SOMP. Combined with the time-varying analog combining matrix, DSOMP and DCOMP have more gain over other techniques with a fixed combining matrix. The gap between DSOMP and DCOMP, however, is marginal in this case. 

 \figref{fig:NB_SingleAnt_64antennas_8RFchains} shows the results with 8 RF chains instead of 16 RF chains. As discussed in \sref{sec:theoretical_analysis}, none of the algorithms using a fixed analog combining matrix such as OMP, SOMP, and COMP  work properly. Moreover,  SOMP  even yields  worse performance than  OMP. 
In contrast,   DSOMP and  DCOMP that use a time-varying sensing matrix have a remarkable gain compared to those that use a fixed sensing matrix. 
In addition,  DCOMP, which exploits the Hermitian property of covariance matrices,  has a considerable gain compared to  DSOMP. These results are consistent with the analysis in \sref{sec:theoretical_analysis}.

\begin{figure}[!t]
	\centering
	\subfigure[center][Efficiency metric vs. $T$]{
		\includegraphics[width=.41\columnwidth]{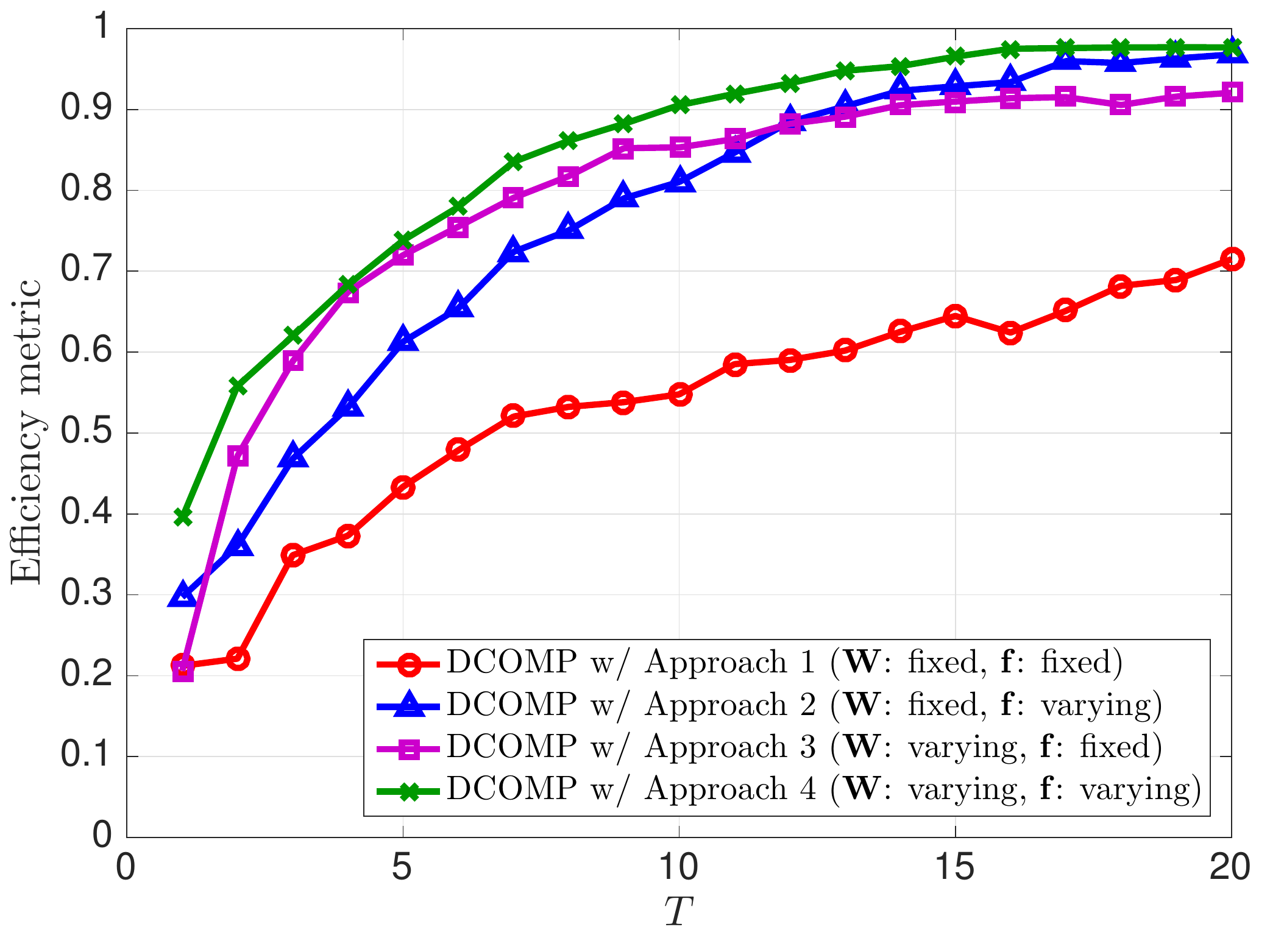}
		\label{fig:NB_MultiAnt_64antennas_effmetric}}
	\subfigure[center][Rate loss vs. $T$]{
		\includegraphics[width=.41\columnwidth]{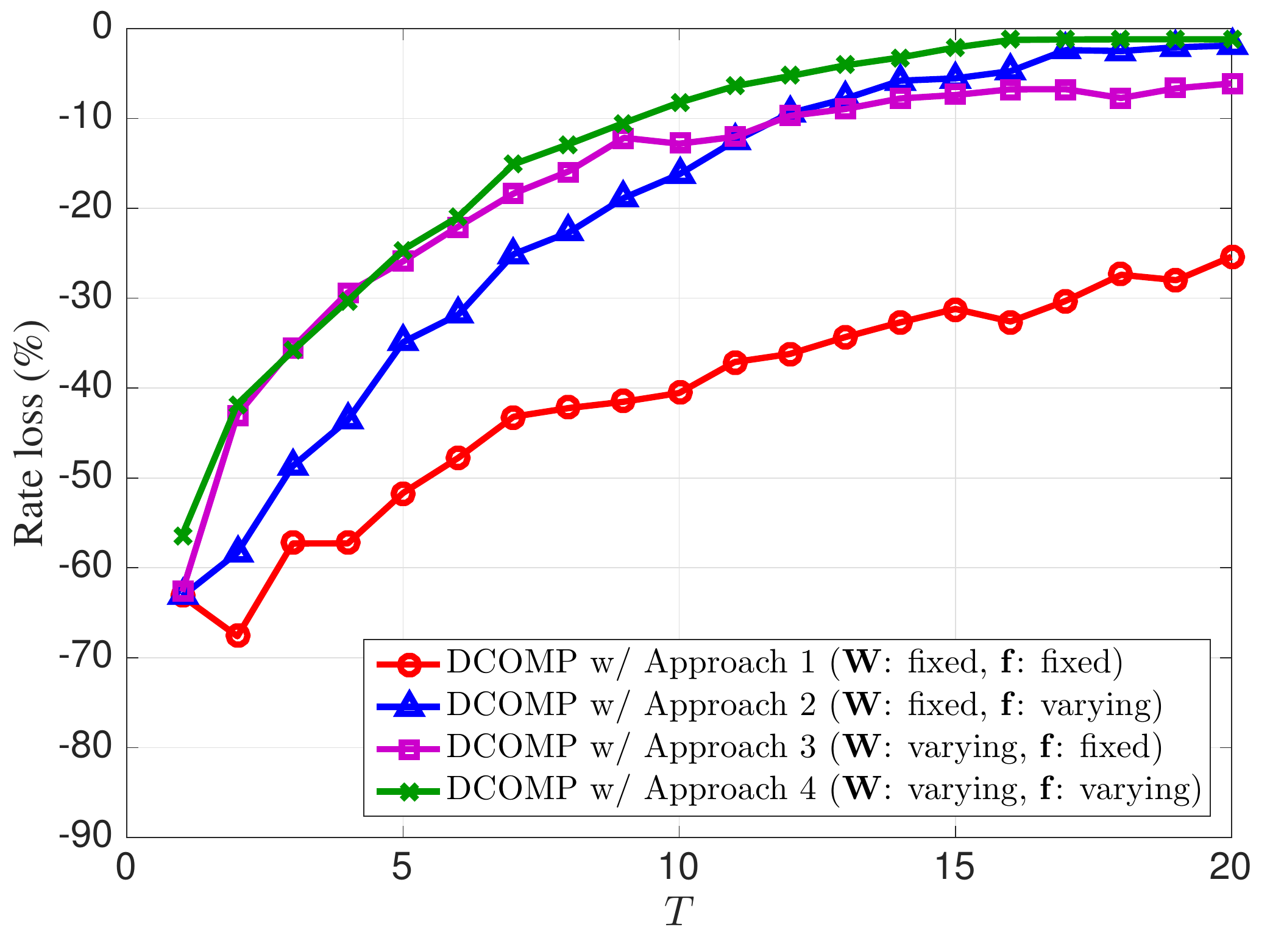}
		\label{fig:NB_MultiAnt_64antennas_rate}}
	\caption{Comparison  among four different approaches when $\NantT = 64$, $\NrfT=8$, $\DT=256$, $\NantR = 64$, $\NrfR=8$, $\DR=256$, $L=8$, and SNR = 0 dB for the multiple-MS-antenna case in narrowband systems.}
	\label{fig:NB_MultiAnt_64antennas}
\end{figure}

\begin{figure}[!t]

	\centering
	\subfigure[center][Efficiency metric vs. $T$]{
		\includegraphics[width=.41\columnwidth]{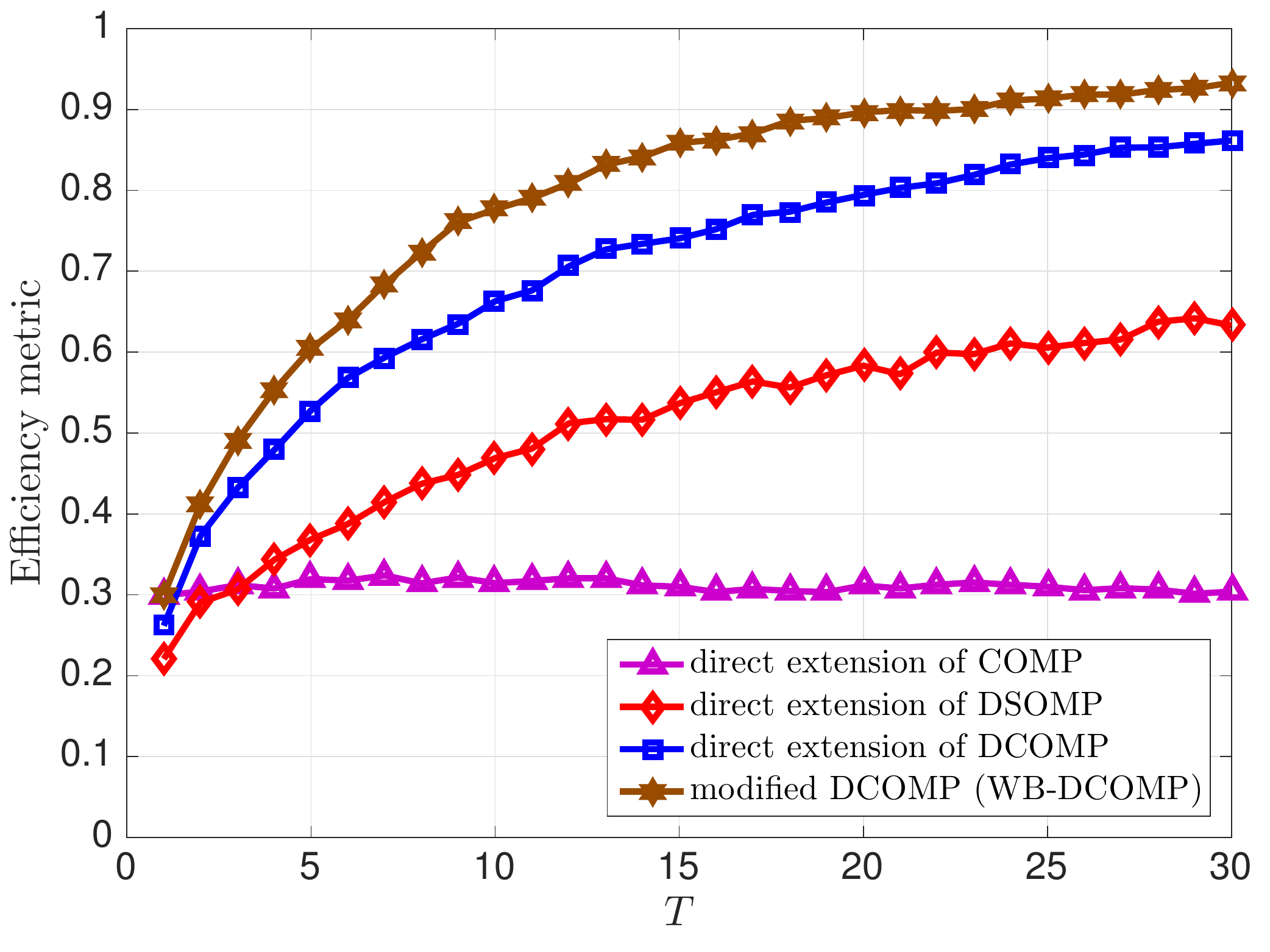}
		\label{fig:WB_SingleAnt_64antennas_effmetric}}
	\subfigure[center][Rate loss vs. $T$]{
		\includegraphics[width=.41\columnwidth]{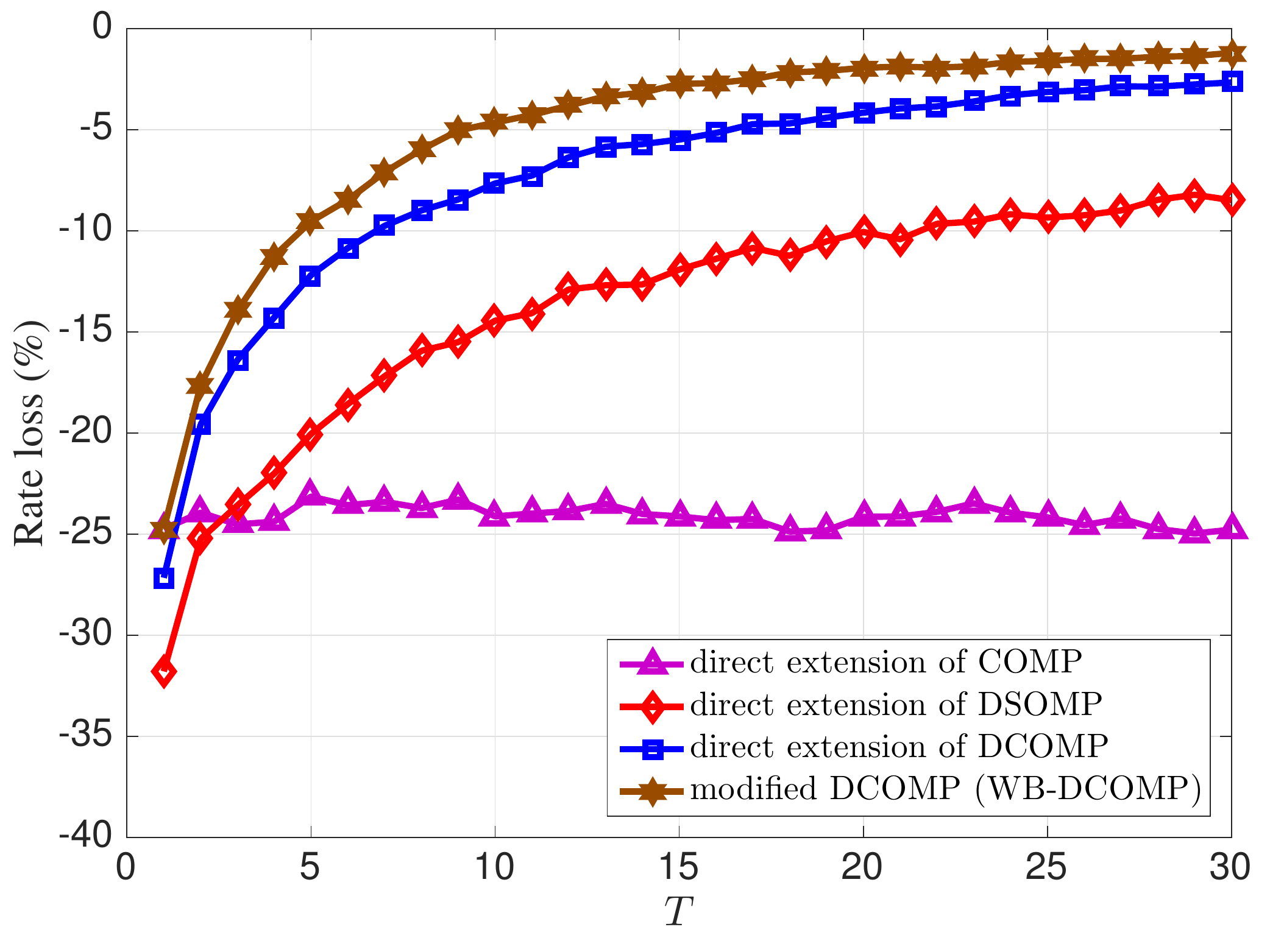}
		\label{fig:WB_SingleAnt_64antennas_rate}}
	\caption{Comparison  among the WB-DCOMP and the direct extensions of other algorithms when $\Nt = 64$, $\Nrf=8$, $D=256$, $L=8$, $K=128$, $N_{\mathrm{CP}}=32$, and SNR = 0 dB for the single-MS-antenna case in wideband systems. The  normalized delays $\tau_{\ell}/T_s$'s are uniformly distributed in $(0, N_{\mathrm{CP}})$, and $p(t)=\mathrm{sinc}(t/T_s)$ is used assuming that ideal low pass filters are employed.
	}
	\label{fig:WB_SingleAnt_64antennas}
\end{figure}

Regarding the extension of the proposed work to the multiple-MS-antenna case,  the four different approaches explained in \sref{subsec:extension_multipleMSant} are compared in  \figref{fig:NB_MultiAnt_64antennas}. As expected,  the use of time-varying precoding and combining matrices at both BS and MS improves covariance estimation. 
\figref{fig:WB_SingleAnt_64antennas} shows the extension to the wideband case. As shown in the figure, the combination of  COMP and  DCOMP outperforms the direct extension of  COMP or  DCOMP. 
In addition to the efficiency metric shown in \figref{fig:NB_MultiAnt_64antennas_effmetric} and \figref{fig:WB_SingleAnt_64antennas_effmetric}, we evaluate the loss  caused by covariance estimation error in terms of spectral efficiency of hybrid precoding.
The rate loss in \figref{fig:NB_MultiAnt_64antennas_rate} and \figref{fig:WB_SingleAnt_64antennas_rate} is defined as $\frac{\mathrm{SE}_{\mathrm{est}} - \mathrm{SE}_{\mathrm{ideal}} }{\mathrm{SE}_{\mathrm{ideal}}} $ in percentage where $\mathrm{SE}_{\mathrm{est}}$ and $\mathrm{SE}_{\mathrm{ideal}}$ are the spectral efficiency of the estimated and ideal covariance case 
under the assumption that the analog precoding matrix is composed of the dominant eigenvectors of the estimated or ideal spatial channel covariance matrix.
The figures show that the trend in the  rate loss is consistent with that in  the efficiency metric although the rate does not only depend on the efficiency metric but also on other factors such as the type of MIMO techniques and the number of streams.

\section{Conclusions}\label{sec:conclusions}
In this paper, we proposed spatial channel covariance estimation techniques for the hybrid MIMO structure.
Based on compressive sensing techniques that leverage the spatially sparse property of the channel, 
we developed covariance estimation algorithms that are featured by two key ideas. 
One is to apply a time-varying sensing matrix, and the other is to exploit the Hermitian property of the covariance matrix. 
Simulation results showed that the proposed greedy algorithms outperform prior work, and the benefit of adopting the two ideas becomes more significant as the number of RF chains becomes smaller. We also analyzed the performance of the proposed algorithms in terms of recovery success probability. 
The theoretical analysis indicated that the success probability approaches one as the number of  snapshots  increases if a time-varying sensing matrix is applied. 
The analysis also proved that using the structured property of the covariance matrix improves the estimation performance, which is consistent with the simulation results.

\appendices
\section{Proof of Theorem 1}\label{app:proof_theorem_1}

Let $ \bP_{t,\cS_o} =  \mathbf{\Phi}_{t,\cS_o} \mathbf{\Phi}_{t,\cS_o}^{\dagger}$. 
Then, the support selection criterion at the $(L_o+1)$-th iteration in the DSOMP shown in \algoref{alg:DSOMP} becomes
\begin{equation}\label{DSOMP_sel_criterion}
\begin{split}
  j^{\mathrm{(opt)}} 
&= \arg \max_{j \in \cN} \sum_{t=1}^{T} \left | \boldsymbol{\phi}_{t,j}^*  \left( \bI - \bP_{t,\cS_o} \right)   \mathbf{\Phi}_{t,\cS} \bg_{t,\cS}  \right|^2 \\
&\stackrel{(a)}{=} \arg \max_{j\in \cN \setminus \cS_o} \sum_{t=1}^{T} \left | \boldsymbol{\phi}_{t,j}^*  \left( \bI - \bP_{t,\cS_o} \right)   \mathbf{\Phi}_{t,\cS \setminus \cS_o} \bg_{t,\cS \setminus \cS_o}  \right|^2 \\ 
&\stackrel{(b)}{=} \arg \max_{j\in \cN \setminus \cS_o} \sum_{t=1}^{T} \left | \sum_{i \in \cS \setminus \cS_o} \boldsymbol{\psi}_{t,j}^* \boldsymbol{\psi}_{t,i} {g}_{t,i} \right|^2 ,
\end{split}
\end{equation}
where $(a)$ comes from $ \mathbf{\Phi}_{t,\cS} \bg_{t,\cS} =  \mathbf{\Phi}_{t,\cS_o} \bg_{t,\cS_o} +  \mathbf{\Phi}_{t,\cS \setminus \cS_o} \bg_{t,\cS \setminus \cS_o}$ and $\left( \bI - \bP_{t,\cS_o}\right)  \mathbf{\Phi}_{t,\cS_o} = \mathbf{0}$, and $(b)$ comes from $( \bI - \bP_{t,\cS_o} )^2 =\bI - \bP_{t,\cS_o}   $ due to the characteristics of the orthogonal projection matrix $\bP_{t,\cS_o}$. Consequently,  one of the elements in the optimal support $\cS$ is selected, i.e., $  j^{\mathrm{(opt)}} \in \cS \setminus \cS_o $ at iteration $L_o+1$ if and only if \eqref{def_rho_DSOMP0} is satisfied.

\section{Proof of Proposition 1}\label{app:proof_prop_1}

If $\mathbf{\Phi}_1 = \cdots = \mathbf{\Phi}_T = \mathbf{\Phi}$, $\rho^{\mathrm{(DS)}}(\cS, \cS_o, \mathbf{\Phi}_1,...,\mathbf{\Phi}_T,\bG)$ in \eqref{def_rho_DSOMP0} becomes
\begin{equation}\label{def_rho_SOMP_proof}
\begin{split}
\rho^{\mathrm{(DS)}}(\cS, \cS_o, \mathbf{\Phi},...,\mathbf{\Phi},\bG)  
 &= \frac{ \max_{j \in \cN \setminus \cS}    \boldsymbol{\psi}_{j}^* \mathbf{\Psi}_{\cS \setminus \cS_o} \left( \frac{1}{T} \sum_{t=1}^{T} \bg_{t} \bg_{t}^* \right) \mathbf{\Psi}_{\cS \setminus \cS_o}^*  \boldsymbol{\psi}_{j} }   {\max_{j \in \cS \setminus \cS_o}  \boldsymbol{\psi}_{j}^* \mathbf{\Psi}_{\cS \setminus \cS_o} \left( \frac{1}{T} \sum_{t=1}^{T} \bg_{t} \bg_{t}^* \right) \mathbf{\Psi}_{\cS \setminus \cS_o}^*  \boldsymbol{\psi}_{j} }   \\
 &\stackrel{(a)}{\rightarrow} \frac{ \max_{j \in \cN \setminus \cS}    \boldsymbol{\psi}_{j}^* \mathbf{\Psi}_{\cS \setminus \cS_o}  \mathbf{\Psi}_{\cS \setminus \cS_o}^*  \boldsymbol{\psi}_{j} }   {\max_{j \in \cS \setminus \cS_o}  \boldsymbol{\psi}_{j}^* \mathbf{\Psi}_{\cS \setminus \cS_o}  \mathbf{\Psi}_{\cS \setminus \cS_o}^*  \boldsymbol{\psi}_{j} }   \\
& = \frac{ \max_{j \in \cN \setminus \cS}   \sum_{i \in \cS \setminus \cS_o} \left |  \boldsymbol{\psi}_{j}^* \boldsymbol{\psi}_{i}   \right |^2 }   {\max_{j \in \cS \setminus \cS_o}  \sum_{i \in \cS \setminus \cS_o} \left |  \boldsymbol{\psi}_{j}^*   \boldsymbol{\psi}_{i}   \right |^2 },
\end{split}
\end{equation} 
where $(a)$ comes from the fact that ${g}_{t,i}$ are independent random variables with zero mean and unit variance.  

\section{Proof of Theorem 2}\label{app:proof_theorem_2}

$\rho^{\mathrm{(DS)}}(\cS,\cS_o,\mathbf{\Phi}_{1},...,\mathbf{\Phi}_{T},\bG)  $ in \eqref{def_rho_DSOMP0} can be written as
\begin{equation}\label{def_rho_UB1}
\begin{split}
\rho^{\mathrm{(DS)}} (\cS,\cS_o,\mathbf{\Phi}_{1},...,\mathbf{\Phi}_{T},\bG)  
&=\frac{ \max_{j \in \cN \setminus \cS}   \frac{1}{T} \sum_{t=1}^T \left( \sum_{i \in \cS \setminus \cS_o}  \left | \boldsymbol{\psi}_{t,j}^* \boldsymbol{\psi}_{t,i}   \right |^2 +   X_{t,j} \right) }   {\max_{j \in \cS \setminus \cS_o}  \frac{1}{T}  \sum_{t=1}^T \left( \sum_{i \in \cS \setminus \cS_o}   \left | \boldsymbol{\psi}_{t,j}^* \boldsymbol{\psi}_{t,i}   \right |^2 + X_{t,j}  \right)  } \\
& \stackrel{(a)}{\rightarrow} \frac{(L - L_o) \mathbb{E}\left[ \left | \boldsymbol{\psi}_{t,j(\neq i)}^* \boldsymbol{\psi}_{t,i}   \right |^2 \right] }{ (L - L_o - 1) \mathbb{E}\left[ \left | \boldsymbol{\psi}_{t,j(\neq i)}^* \boldsymbol{\psi}_{t,i}   \right |^2 \right]  +  \mathbb{E}\left[ \left | \boldsymbol{\psi}_{t,i}^* \boldsymbol{\psi}_{t,i}   \right |^2 \right] },
\end{split}
\end{equation}
where $X_{t,j} =   \sum_{i_1 \in \cS \setminus \cS_o} \sum_{\substack{ i_2 \in \cS \setminus \cS_o \\ i_2 \neq  i_1 } }  {g}_{t,i_1} {g}_{t,i_2}^* \boldsymbol{\psi}_{t,j}^* \boldsymbol{\psi}_{t,i_1} \boldsymbol{\psi}_{t,i_2}^* \boldsymbol{\psi}_{t,j}$. In \eqref{def_rho_UB1}, 
 $(a)$ comes from the fact that $\boldsymbol{\psi}_{t,i}^* \boldsymbol{\psi}_{t,i}$ for all $i$ are identical random variables with non-zero mean,  $\boldsymbol{\psi}_{t,j}^* \boldsymbol{\psi}_{t,i}$ for all $j \neq i$ are  identical random variables with zero-mean, and $g_{t,i}$ are independent random variables with zero mean and unit variance. 

Now let us look at  $ \mathbb{E}\left[ \left | \boldsymbol{\psi}_{t,i}^* \boldsymbol{\psi}_{t,i}   \right |^2 \right] $ and $\mathbb{E}\left[ \left | \boldsymbol{\psi}_{t,j(\neq i)}^* \boldsymbol{\psi}_{t,i}   \right |^2 \right]$. 
Let $\mathbf{\Omega}_{t} = \mathbf{\Psi}_{t,\cN \setminus \cS_o}^* \mathbf{\Psi}_{t,\cN \setminus \cS_o}$, then 
the trace of $\mathbf{\Omega}_{t}$ becomes
\begin{equation}\label{traceG}
\begin{split}
\mathrm{Tr}(\mathbf{\Omega}_{t}) &=  \mathrm{Tr} \left(  \mathbf{\Phi}_{t,\cN \setminus \cS_o} ^* \left( \bI_{\Nrf} - \bP_{t,\cS_o} \right)^2 \mathbf{\Phi}_{t,\cN \setminus \cS_o}  \right) \\
& \stackrel{(a)}{=} \mathrm{Tr} \left(   \left( \bI_{\Nrf} - \bP_{t,\cS_o} \right) \left( \Nt \bI_{\Nrf} - \mathbf{\Phi}_{t, \cS_o} \mathbf{\Phi}_{t, \cS_o} ^* \right)  \right) \\
& \stackrel{(b)}{=} \Nt \left( \Nrf - L_o \right),
\end{split}
\end{equation}
where $(a)$ comes from $\mathbf{\Phi}_{t,\cS_o} \mathbf{\Phi}_{t,\cS_o} ^* +  \mathbf{\Phi}_{t,\cN \setminus \cS_o}  \mathbf{\Phi}_{t,\cN \setminus \cS_o} ^* = \Nt \bI_{\Nt}$ due to the tight frame property, and $(b)$ comes from $\left( \bI - \bP_{t,\cS_o}\right)  \mathbf{\Phi}_{t,\cS_o} = \mathbf{0}$.
From \eqref{traceG}, the lower bound of $ \mathbb{E}\left[ \left | \boldsymbol{\psi}_{t,i}^* \boldsymbol{\psi}_{t,i}   \right |^2 \right] $ is given by
\begin{equation}\label{E_ii_abs2_LB}
\begin{split}
\mathbb{E}\left[ \left | \boldsymbol{\psi}_{t,i}^* \boldsymbol{\psi}_{t,i}   \right |^2 \right]  & \stackrel{(a)}{\geq} \left( \mathbb{E}\left[ \left | \boldsymbol{\psi}_{t,i}^* \boldsymbol{\psi}_{t,i}   \right | \right] \right)^2  
= \left( \frac{\mathbb{E}\left[\mathrm{Tr}(\mathbf{\Omega}_{t}) \right]}{\Nt - L_o } \right)^2 
=\frac{\Nt^2 (\Nrf - L_o)^2}{(\Nt - L_o)^2},
\end{split}
\end{equation}
where $(a)$ comes from the fact that $\mathbb{E}[|X|^2] \geq \left(\mathbb{E}[|X|] \right)^2$ for any random variable $X$.

Now, let us look at the squared Frobenius norm of $\mathbf{\Omega}_{t}$ that is given by
\begin{equation}\label{FrobNormG}
\begin{split}
\|\mathbf{\Omega}_{t} \|^2_{\mathrm{F}} 
&= \mathrm{Tr} \left( \left( \left( \bI_{\Nrf} - \bP_{t,\cS_o} \right) \mathbf{\Phi}_{t,\cN \setminus \cS_o}  \mathbf{\Phi}_{t,\cN \setminus \cS_o} ^* \right)^2 \right) \\
&= \Nt^2 \left(\Nrf - L_o \right).
\end{split}
\end{equation}
In addition, $\mathbb{E} \left[ \| \mathbf{\Omega}_t \|^2_{\mathrm{F}} \right] $ can be represented differently as
\begin{equation}\label{Frob_Gram_1}
\begin{split}
\mathbb{E} \left[ \| \mathbf{\Omega}_t \|^2_{\mathrm{F}} \right] 
&=\mathbb{E} \left[ \sum_{i \in \cN \setminus \cS_o}  \left | \boldsymbol{\psi}_{t,i}^* \boldsymbol{\psi}_{t,i}   \right |^2  + \sum_{\substack{ j,i \in \cN \setminus \cS_o \\ j \neq i}}  \left | \boldsymbol{\psi}_{t,j}^* \boldsymbol{\psi}_{t,i}   \right |^2  \right] \\
&= (\Nt - L_o) \mathbb{E} \left[ \left | \boldsymbol{\psi}_{t,i}^* \boldsymbol{\psi}_{t,i}   \right |^2 \right] + (\Nt - L_o)(\Nt - L_o-1) \mathbb{E} \left[ \left | \boldsymbol{\psi}_{t,j(\neq i)}^* \boldsymbol{\psi}_{t,i}   \right |^2  \right]. 
\end{split}
\end{equation}
From \eqref{E_ii_abs2_LB}-\eqref{Frob_Gram_1},  the upper bound of  $\mathbb{E}\left[ \left | \boldsymbol{\psi}_{t,j(\neq i)}^* \boldsymbol{\psi}_{t,i}   \right |^2 \right]$ can be obtained  as  
\begin{equation}\label{E_ji_abs^2_UB}
\begin{split}
\mathbb{E}\left[ \left | \boldsymbol{\psi}_{t,j(\neq i)}^* \boldsymbol{\psi}_{t,i}   \right |^2 \right] 
&=  \frac{ \Nt^2 \left(\Nrf - L_o \right) - (\Nt - L_o) \mathbb{E} \left[ \left | \boldsymbol{\psi}_{t,i}^* \boldsymbol{\psi}_{t,i}   \right |^2 \right] }{(\Nt - L_o)(\Nt - L_o-1) } \\
& \leq \frac{\Nt^2 (\Nrf - L_o) (\Nt - \Nrf)}{(\Nt - L_o)^2 (\Nt - L_o - 1)} .
\end{split}
\end{equation}

By using the lower bound of $ \mathbb{E}\left[ \left | \boldsymbol{\psi}_{t,i}^* \boldsymbol{\psi}_{t,i}   \right |^2 \right] $ and the upper bound of $\mathbb{E}\left[ \left | \boldsymbol{\psi}_{t,j(\neq i)}^* \boldsymbol{\psi}_{t,i}   \right |^2 \right]$, it can be shown that the converged value of $\rho^{\mathrm{(DS)}}(\cS,\cS_o,\mathbf{\Phi}_{1},...,\mathbf{\Phi}_{T}, \bG)$  in \eqref{def_rho_UB1} has an upper bound as
\begin{equation}\label{def_rho_UB2}
\begin{split}
\rho ^{\mathrm{(DS)}} (\cS,\cS_o,\mathbf{\Phi}_{1},...,\mathbf{\Phi}_{T}, \bG)  
& \rightarrow \frac{L - L_o  }{ L - L_o - 1  + \frac{ \mathbb{E}\left[ \left | \boldsymbol{\psi}_{t,i}^* \boldsymbol{\psi}_{t,i}   \right |^2 \right]}{\mathbb{E}\left[ \left | \boldsymbol{\psi}_{t,j(\neq i)}^* \boldsymbol{\psi}_{t,i}   \right |^2 \right]} } \\
& \leq  \frac{L - L_o  }{ L - L_o - 1  +   \frac{ (\Nrf - L_o) (\Nt - L_o - 1)}{ (\Nt - \Nrf)}},\end{split}
\end{equation}
and this upper bound is always less than or equal to one because $\Nt \geq \Nrf \geq L > L_o$.

\section{Proof of Theorem 4}\label{app:proof_theorem_4}

As $T \rightarrow \infty$, $\rho^{\mathrm{(DS)}}(\cS, \cS_o, \mathbf{\Phi}_1,...,\mathbf{\Phi}_T, \bG)$  in \eqref{def_rho_DSOMP0} can be rewritten as
\begin{equation}\label{def_rho_DSOMP_re}
\begin{split}
\rho^{\mathrm{(DS)}}(\cS, \cS_o, \mathbf{\Phi}_1,...,\mathbf{\Phi}_T, \bG)  
 &= \frac{ \max_{j \in \cN \setminus \cS}   \sum_{t=1}^{T} \boldsymbol{\phi}_{t,j}^*   \bQ_{t,\cS,\cS_o}^{\mathrm{(DS)}}  \boldsymbol{\phi}_{t,j}}   {\max_{j \in \cS \setminus \cS_o}  \sum_{t=1}^{T} \boldsymbol{\phi}_{t,j}^*   \bQ_{t,\cS,\cS_o}^{\mathrm{(DS)}}  \boldsymbol{\phi}_{t,j} } \\
& \stackrel{(a)}{\rightarrow} \frac{\frac{1}{\Nt-L}\mathbb{E} \left[ \mathrm{Tr}(\mathbf{\Phi}_{t,\cN \setminus \cS}^* \tilde{\bQ}_{t,\cS,\cS_o}^{\mathrm{(DS)}}  \mathbf{\Phi}_{t,\cN \setminus \cS} ) \right] }   {\frac{1}{L-L_o}\mathbb{E}\left[ \mathrm{Tr}(\mathbf{\Phi}_{t,\cS \setminus \cS_o}^* \tilde{\bQ}_{t,\cS,\cS_o}^{\mathrm{(DS)}}  \mathbf{\Phi}_{t,\cS \setminus \cS_o} )  \right]}   ,  
\end{split}
\end{equation}
where $\tilde{\bQ}_{t,\cS,\cS_o}^{\mathrm{(DS)}}  = \left( \bI_M - \bP_{t,\cS_o} \right)   \mathbf{\Phi}_{t,\cS \setminus \cS_o}   \mathbf{\Phi}_{t,\cS \setminus \cS_o}^* \left( \bI_M - \bP_{t,\cS_o} \right)$
and $(a)$ comes from the fact that $\bG$ and $\mathbf{\Phi}_t$ are independent and all elements in $\bG$ have zero mean and unit variance.

In a similar way, $\rho^{\mathrm{(DC)}}(\cS, \cS_o, \mathbf{\Phi}_1,...,\mathbf{\Phi}_T, \bG)$  in \eqref{def_rho_DCOMP} converges as
\begin{equation}\label{def_rho_DCOMP_re}
\begin{split}
\rho^{\mathrm{(DC)}}(\cS, \cS_o, \mathbf{\Phi}_1,...,\mathbf{\Phi}_T, \bG)   
& \rightarrow \frac{\frac{1}{\Nt-L}\mathbb{E} \left[ \mathrm{Tr}(\mathbf{\Phi}_{t,\cN \setminus \cS}^* \tilde{\bQ}_{t,\cS,\cS_o}^{\mathrm{(DC)}}  \mathbf{\Phi}_{t,\cN \setminus \cS} ) \right] }   {\frac{1}{L-L_o}\mathbb{E}\left[ \mathrm{Tr}(\mathbf{\Phi}_{t,\cS \setminus \cS_o}^* \tilde{\bQ}_{t,\cS,\cS_o}^{\mathrm{(DC)}}  \mathbf{\Phi}_{t,\cS \setminus \cS_o} )  \right]}   ,
\end{split}
\end{equation}
where 
$\tilde{\bQ}_{t,\cS,\cS_o}^{\mathrm{(DC)}} =   \mathbf{\Phi}_{t,\cS \setminus \cS_o} ^* \mathbf{\Phi}_{t,\cS \setminus \cS_o}^*  - \bP_{t,\cS_o} \mathbf{\Phi}_{t,\cS \setminus \cS_o}   \mathbf{\Phi}_{t,\cS \setminus \cS_o}^* \bP_{t,\cS_o}$.

Since the terms in the expectation in \eqref{def_rho_DSOMP_re} and \eqref{def_rho_DCOMP_re} are independent random variables with respect to $t$, we omit the time slot index $t$  for simplicity.  The difference between the inner parts of the denominators in \eqref{def_rho_DSOMP_re} and \eqref{def_rho_DCOMP_re} becomes 
\begin{equation}\label{Q_diff_in1}
\begin{split}
 \mathrm{Tr} \hspace{-2pt} \left( \mathbf{\Phi}_{\cS \setminus \cS_o}^* \left(\tilde{\bQ}_{\cS,\cS_o}^{\mathrm{(DC)}}  - \tilde{\bQ}_{\cS,\cS_o}^{\mathrm{(DS)}} \right) \mathbf{\Phi}_{\cS \setminus \cS_o}  \right)  & = 2 \mathrm{Re} \hspace{-2pt} \left( \mathrm{Tr}  \left(  \mathbf{\Phi}_{\cS \setminus \cS_o}^*  ( \bI  - \bP_{\cS_o}  ) \mathbf{\Phi}_{\cS \setminus \cS_o}   \mathbf{\Phi}_{\cS \setminus \cS_o}^* \bP_{\cS_o} \mathbf{\Phi}_{\cS \setminus \cS_o}  \right)  \right) \hspace{-2pt}. \\  
\end{split}
\end{equation}
Note that both $\bP_{\cS_o}$ and $\bI - \bP_{\cS_o}$ can be represented as the production of two semi-unitary matrices, and thus $\mathbf{\Phi}_{\cS \setminus \cS_o}^*  \bP_{\cS_o} \mathbf{\Phi}_{\cS \setminus \cS_o} $ and $\mathbf{\Phi}_{\cS \setminus \cS_o}^*  \left( \bI - \bP_{\cS_o}  \right) \mathbf{\Phi}_{\cS \setminus \cS_o} $  become positive semidefinite matrices. 
Since the trace of the product of two semidefinite matrices is larger than or equal to zero,
 \eqref{Q_diff_in1} becomes
\begin{equation}\label{Q_diff_in2}
\begin{split}
 \mathrm{Tr} \hspace{-2pt} \left( \mathbf{\Phi}_{\cS \setminus \cS_o}^* \left(\tilde{\bQ}_{\cS,\cS_o}^{\mathrm{(DC)}}  - \tilde{\bQ}_{\cS,\cS_o}^{\mathrm{(DS)}} \right) \mathbf{\Phi}_{\cS \setminus \cS_o}  \right)  \geq 0 
\end{split}
\end{equation}
for any $\mathbf{\Phi}$, $\cS$, and $\cS_o$. 
The difference between the inner parts of the numerators in \eqref{def_rho_DSOMP_re} and \eqref{def_rho_DCOMP_re} becomes
\begin{equation}\label{Q_diff_out1_1}
\begin{split}
 \hspace{-3pt} \mathrm{Tr} \hspace{-2pt} \left( \mathbf{\Phi}_{\cN \setminus \cS}^* \hspace{-2pt} \left(\tilde{\bQ}_{\cS,\cS_o}^{\mathrm{(DC)}}  - \tilde{\bQ}_{\cS,\cS_o}^{\mathrm{(DS)}} \right) \hspace{-2pt}  \mathbf{\Phi}_{\cN \setminus \cS}  \right)  
&=  2 \mathrm{Re} \hspace{-2pt} \left(\mathrm{Tr} \hspace{-2pt}\left(  \mathbf{\Phi}_{\cN \setminus \cS}^*  \left( \bI  - \bP_{\cS_o}  \right) \mathbf{\Phi}_{\cS \setminus \cS_o}   \mathbf{\Phi}_{\cS \setminus \cS_o}^* \bP_{\cS_o} \mathbf{\Phi}_{\cN \setminus \cS}  \right)\right)  \\
&\stackrel{(a)}{=} -  2 \mathrm{Re} \hspace{-2pt} \left(\mathrm{Tr} \hspace{-2pt} \left(   \bP_{\cS_o} \mathbf{\Phi}_{\cS \setminus \cS_o}  \mathbf{\Phi}_{\cS \setminus \cS_o}^*  \left( \bI - \bP_{\cS_o}  \right) \mathbf{\Phi}_{\cS \setminus \cS_o}  \mathbf{\Phi}_{\cS \setminus \cS_o}^*     \right) \right)\\
& = -  \mathrm{Tr} \hspace{-2pt} \left( \mathbf{\Phi}_{\cS \setminus \cS_o}^* \left(\tilde{\bQ}_{\cS,\cS_o}^{\mathrm{(DC)}}  - \tilde{\bQ}_{\cS,\cS_o}^{\mathrm{(DS)}} \right) \mathbf{\Phi}_{\cS \setminus \cS_o}  \right) \\
 &\leq 0  ,
\end{split}
\end{equation}
where $(a)$ can be proved by using  $\mathbf{\Phi}_{\cN \setminus \cS}  \mathbf{\Phi}_{\cN \setminus \cS}^* = N \bI_M -   \mathbf{\Phi}_{\cS_o}  \mathbf{\Phi}_{\cS_o}^* -  \mathbf{\Phi}_{\cS \setminus \cS_o}  \mathbf{\Phi}_{\cS \setminus \cS_o}^* $ 
and  $ \bP_{\cS_o} ( N \bI_M -   \mathbf{\Phi}_{\cS_o}  \mathbf{\Phi}_{\cS_o}^*  ) ( \bI_M - \bP_{\cS_o}  ) = \mathbf{0}$.

From the two inequalities in \eqref{Q_diff_in2} and \eqref{Q_diff_out1_1}, the converged value in \eqref{def_rho_DSOMP_re} is always larger than or equal to that  in \eqref{def_rho_DCOMP_re}, and this completes the proof.

\nobalance

\bibliographystyle{IEEEtran}
\bibliography{IEEEabrv,Bib_cov_est}

\end{document}